\documentclass[twocolumn, superscriptaddress, nofootinbib, longbibliography]{revtex4-1} 

\usepackage{amsmath,amssymb}
\usepackage{amsfonts}
\usepackage{braket}
\usepackage{slashed}	
\usepackage[export]{adjustbox}
\usepackage{graphicx}
\usepackage{fancybox,xcolor}
\usepackage{float}
\usepackage[font=footnotesize,labelfont=bf, justification=centerlast]{caption}
\usepackage[normalem]{ulem}
\usepackage{xspace}
\usepackage{enumitem}
\setitemize{topsep=0pt,itemsep=2pt,partopsep=0pt, parsep=0pt}
\setenumerate{noitemsep,topsep=2pt,parsep=2pt,partopsep=2pt}

\usepackage[utf8]{inputenc} 
\usepackage[T1]{fontenc} 
\usepackage[english]{babel}
\usepackage{lmodern}

\usepackage{amsthm}
\theoremstyle{remark}

\def\ra{\rightarrow}

\def\lra{\leftrightarrow}

\def\tbul{\noindent \hskip 1em $\bullet$ }

\def\hc{\text{h.c.}}
\def\Heis{\text{Heis}}

\def\1{\mathbf{1}}
\renewcommand{\Re}{\mbox{Re}\,}

\def\O#1{{\mathcal O}\left(#1\right)}
\def\Ocal#1{{\mathcal O}\left(#1\right)}
\def\Tr{\mathrm{Tr}}

\def\L{{\cal L}}
\renewcommand{\ell}{L}
\def\M{{\cal M}}

\def\hc{{\rm{h.c.}}} 
\def\>{\rangle}
\def\<{\langle}
\newcommand{\ev}[1]{\Braket{#1}}
\newcommand{\ketbra}[2]{\Ket{#1}\Bra{#2}}
\def\v{\mathbf} 
\def\p{\partial}

\def\t{\widetilde}

\def\s{{}^\dagger}

\def\eps{\varepsilon}
\def\al{\alpha}
\def\be{\beta}
\def\d{\delta}
\def\D{\Delta}

\def\ka{\kappa}
\def\si{\sigma}
\def\Si{\Sigma}
\def\om{\omega}
\def\ga{\gamma}
\def\Ga{\Gamma}
\def\su{\uparrow}
\def\sd{\downarrow}
\def\NDH{N_\text{DH}}
\def\Okin{O_\text{kin}}
\def\EF{\mu} 

\AtBeginDocument{%
    \newwrite\bibnotes
    \def\bibnotesext{Notes.bib}
    \immediate\openout\bibnotes=\jobname\bibnotesext
    \immediate\write\bibnotes{@CONTROL{REVTEX41Control}}
    \immediate\write\bibnotes{@CONTROL{%
    apsrev41Control,author="08",editor="1",pages="1",title="0",year="1"}}
     \if@filesw
     \immediate\write\@auxout{\string\citation{apsrev41Control}}%
    \fi
}%

\begin{document}

\title{Relaxation to persistent currents in a Hubbard trimer coupled to fermionic baths}

\author{Nikodem Szpak}
\email{nikodem.szpak@uni-due.de}
\affiliation{Fakult\"at f\"ur Physik and CENIDE, Universit\"at Duisburg-Essen, Lotharstra{\ss}e 1, 47057 Duisburg, Germany}
\author{Gernot Schaller}
\affiliation{Helmholtz-Zentrum Dresden-Rossendorf, Bautzner Landstra{\ss}e 400, 01328 Dresden, Germany}
\author{Ralf Sch\"utzhold}
\affiliation{Helmholtz-Zentrum Dresden-Rossendorf, Bautzner Landstra{\ss}e 400, 01328 Dresden, Germany}
\affiliation{Institut f\"ur Theoretische Physik, Technische Universit\"at Dresden, 01062 Dresden, Germany}
\author{J\"urgen K\"onig}
\affiliation{Fakult\"at f\"ur Physik and CENIDE, Universit\"at Duisburg-Essen, Lotharstra{\ss}e 1, 47057 Duisburg, Germany}
 
\date{\today}

\begin{abstract}
  We consider a ring of fermionic quantum sites, modeled by the Fermi--Hubbard Hamiltonian, in which electrons can move and interact strongly via the Coulomb repulsion.
  The system is coupled to fermionic cold baths which by the exchange of particles and energy induce relaxation in the system.
  We eliminate the environment and describe the system effectively by Lindblad master equations in various versions valid
  for different coupling parameter regimes.
  The early relaxation phase  proceeds in a universal way, irrespective of the relative couplings and approximations.
  The system settles down to its low--energy sector and is consecutively well approximated by the Heisenberg model.
  {We compare different Lindblad approaches which}, in the late relaxation, push the system towards different final states with opposite, extreme spin orders, from ferromagenetic to antiferromagnetic.
  Due to spin frustration in the trimer (a three site ring),
  degenerate ground states are formed by spin waves (magnons).
  The system described by the global coherent version of the Lindblad operators relaxes towards the final states carrying directed persistent spin currents.
  We numerically confirm these predictions.
\end{abstract}

\maketitle

\section{Introduction} 

It is a long standing problem how to accurately and at the same time efficiently describe a quantum system coupled to an infinite environment \cite{Breuer, weiss1993, mandel1995}. Each attempt encounters the difficulty of dealing with a tremendous number of degrees of freedom in a huge Hilbert space and the practical impossibility of solving the equations of evolution exactly as soon as any interactions are in play \cite{meir1992a, queisser2019a, kolovsky2020a, nakagawa2021a}.
Prominent examples are atoms, molecules, nanostructures or qubits in real environments \cite{weiss1993, Gardiner-QuantumNoise}. 
Quantum dots coupled to bosonic or fermionic baths belong to the simplest systems of this kind \cite{QDots-Review, QDots+reservoir} but still their proper description is far from being satisfactory and complete.
Although the dynamics of the quantum dots themselves can be solved by various methods exactly, analytically or numerically, their coupling to an infinite environment poses a serious difficulty.
It can change the dynamics from simple oscillatory modes to highly complex relaxation processes going through various stages \cite{leggett1987a}.
In order to achieve an effective approximation scheme, the environment's degrees of freedom must be effectively eliminated from the description {which} requires some simplifying assumptions about their behavior. 
To the most common belong the Born--Markov assumptions leading to the Redfield equation 
\cite{redfield1965a} together with the local {or} global secular approximations 
\cite{farina2019a, maekelae2013a, hofer2017a} reducing it further to the positivity preserving Lindblad (or Gorini--Kossakowski--Sudarshan--Lindblad, GKSL) equation \cite{lindblad1976a, gorini1976a}.

Because of the known limitations in the description of quantum coherences within the system, several improvements of these methods have been proposed 
\cite{MarkovMemory+Slippage, StayingPositive, schultz2009a, trushechkin2021a, Hatrmann+Strunz-RedfieldNonpositivity, Potts2021, Becker+Eckardt-LindBeyondWeak, Abbruzzo-TimedepRegRedfield}.
In \cite{B07-Dimer} we worked out a refinement towards including the maximum amount of coherence allowed mathematically, along similar lines as postulated in \cite{Breuer-CohLind, Kirsanskas+Wacker-CohLindPhenom},
later discussed in \cite{Davidovic2020} 
and formally derived in \cite{Nathan-CohLindDerivation}, which we develop here further. 
In order to describe the dynamics and relaxation processes taking place in general Fermi--Hubbard systems coupled to fermionic baths we developed the method of \textit{local} and \textit{global coherent} Lindblad operators (dissipators) which eliminate the environmental degrees of freedom and offer an effective description of the evolution of the reduced system while keeping the maximal information about the relevant coherences in the system. In this respect, this approach is superior to the standard \textit{secular} approximation and to \textit{local} dissipators which can be obtained from it in some limiting cases.
For the demonstration of the main differences with the \textit{local} Lindblads as well as with the \textit{secular} (incoherent) approach, we applied the method to the Hubbard dimer system \cite{B07-Dimer}. 
Here, we want to consider larger systems and demonstrate that some non--trivial final states of evolution can be obtained when using the \textit{global coherent} approach.

\begin{figure}[ht]
  \begin{center}
    \includegraphics[width=0.9\linewidth]{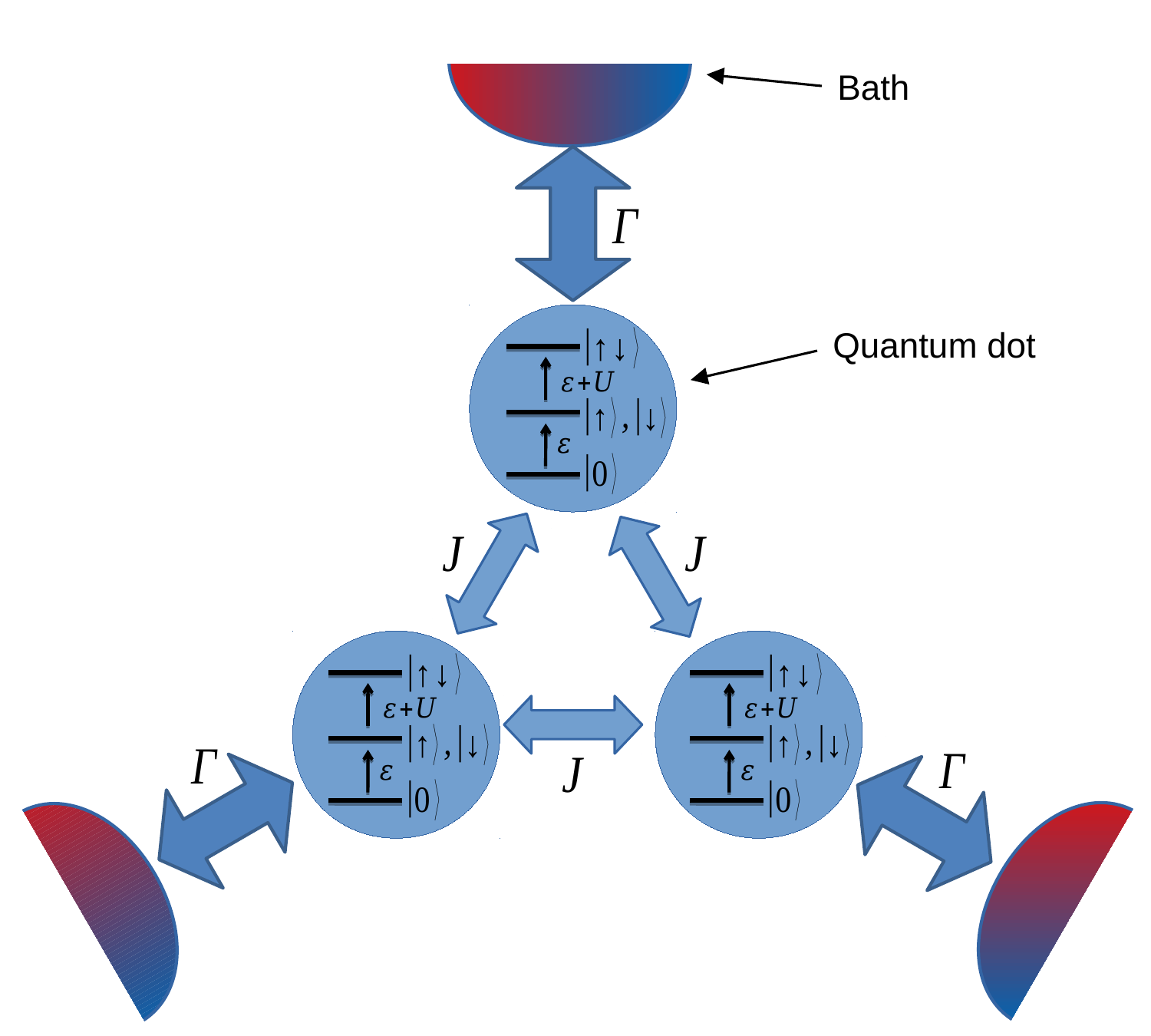}
  \end{center}
  \caption{Schematic presentation of a Hubbard trimer realized by three quantum dots in a ring coupled to the environment. Each dot has its own structure of local energy states and all dots are coupled to their neighbors ($J$) and to local baths ($\Gamma$).
  \label{fig:Trimer}}
\end{figure}

As a specific example, we consider the Hubbard trimer, i.e. three quantum sites ($M=3$), connected periodically in the form of a triangle (or ring), cf. Fig.~\ref{fig:Trimer}.
This is the smallest Hubbard system which shows spin frustration at low energies when driven to half filling by strong repulsive on--site interaction ($U$), large negative binding on--site energy ($\eps$) and low bath temperature ($T$).
We will demonstrate that, within the global coherent approach, valid in the regime of weak coupling to the environment ($\Ga$) relative to the internal couplings ($J$), the system will behave anti--ferromagnetically and settle down to magnonic states carrying persistent spin currents \cite{QuantumRingsForBeginners, PersCur, Kopietz-PersSpinCur, PersSpinCur, PersCurDiracRings}.
In contrast, in the local coherent approach, valid in the regime $J \lesssim \Ga$ {\cite{B07-Tetramer, GSchaller-Review-NQSys}}, the system at half filling will increase its spin order and become ferromagnetic.

\begin{figure}[ht]
  \begin{center}
    \includegraphics[width=\linewidth,trim=20 380 20 40, clip]{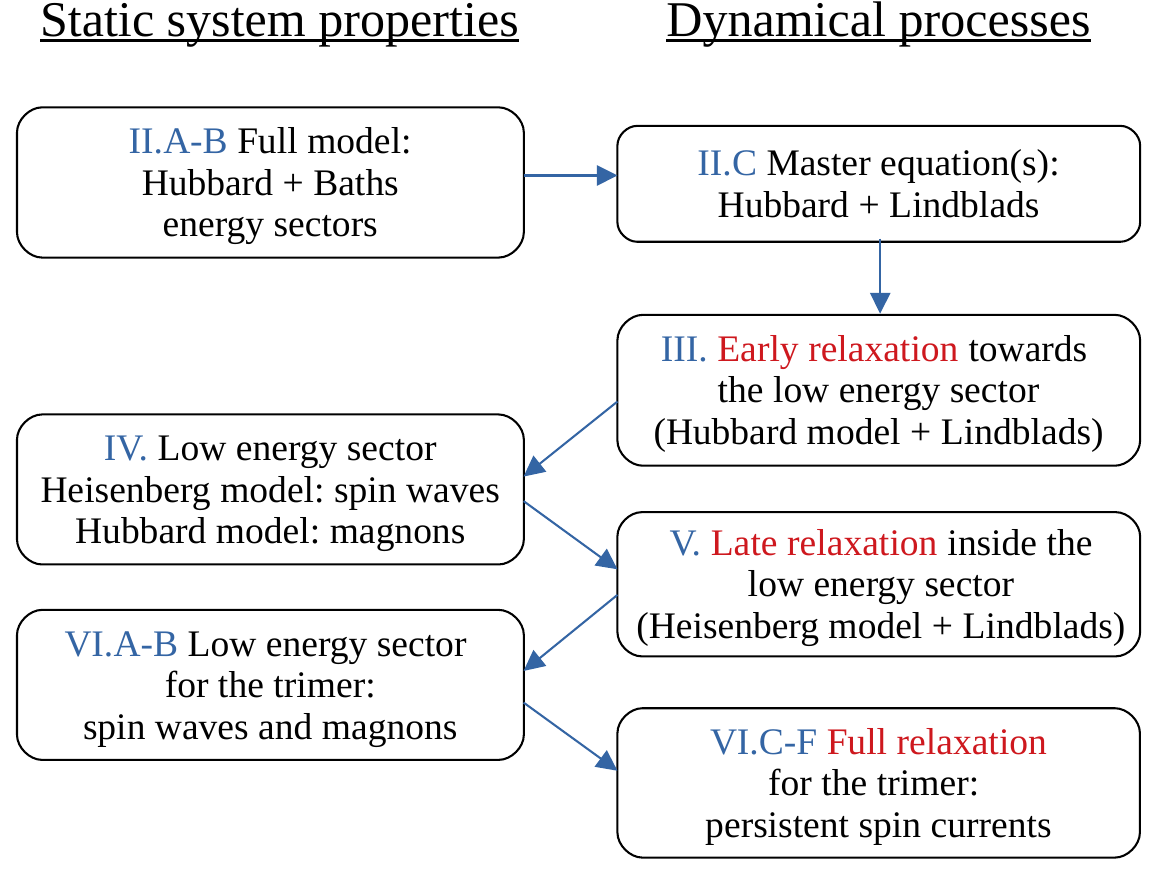}
  \end{center}
  \caption{Schematic presentation of the article's structure, illustrating the logical connections between the sections. 
  \label{fig:structure}}
\end{figure}

This article is organized as follows (cf. Fig.~\ref{fig:structure}). 
In Sec.~\ref{sec:System} we introduce the ring of quantum sites coupled to fermionic cold baths. The exchange of particles and energy with the environment will induce the relaxation in the system. 
Further,
we eliminate the baths from the description and derive effective Lindblad master equations in various versions. The particular Lindblad operators will depend on the used approximations.
We also analyze the energy spectrum of the isolated system ($\Ga = 0$) and distinguish high and low energy sectors (for $U\gg J$). We recognize the low--energy sector to be half filled, with 
doublon and holon contributions suppressed by $J/U$. 
In Sec.~\ref{sec:Early-Relax} we show how the Lindblad approaches universally, i.e. independent of the relative coupling strengths, trigger the relaxation of the system towards the low--energy sector.
In Sec.~\ref{sec:Low-Energy} we study the properties of the low--energy sector where the Hubbard Hamiltonian is approximated by the Heisenberg model with exactly one electron per site. Its energy spectrum is contained in a narrow band and the states can be classified according to the total spin-$z$ component and the spin wavenumber from which the basis of spin waves emerges. 
We then return to the full Hubbard model and, guided by the observations from the Heisenberg model, identify the spin wave states as the lowest energetic states known as magnons.
In Sec.~\ref{sec:Late-Relax} we discuss how various Lindblad operators drive the effective Heisenberg system towards various final states with different spin order. 
In Sec.~\ref{sec:Trimer} we concentrate on the trimer ($M=3$) and discover that, within the coherent approximation, it settles down to magnonic modes carrying persistent spin current.
We provide numerical evidence for the above described relaxation schemes and study it under different conditions, at zero or non--zero temperature and with additional magnetic flux which lifts the ground state degeneracy.

\section{Full model and master equations} \label{sec:System} 
\subsection{Hubbard system and the environment}
\def\sumint{\mathop{\sum\kern-3ex\int}}

We consider a periodic chain, also known as a quantum ring, of $M$ sites connected by tunneling with amplitude $J$ to their nearest neighbors, cf. Fig.~\ref{fig:Trimer}. Each site contains one bound state with energy $\eps<0$ which can be occupied by up to two electrons with opposite spins that interact by the on--site Coulomb repulsion of strength $U>0$. 
The total Fermi--Hubbard Hamiltonian of the system reads
\begin{equation} \label{Hubbard}
  H_S = - J \sum_{\<m,n\>, s} c\s_{m,s} c_{n,s}  + \eps \sum_{m,s} n_{m,s}
  + U \sum_m n_{m,\sd} n_{m,\su} 
\end{equation}
where $c\s_{m,s}$ and  $c_{m,s}$ are the fermionic creation and annihilation operators and
$n_{m,s} = c\s_{m,s} c_{m,s}$ are the particle number operators for electrons at site $m \in \{1, ..., M\}$ with spin $s \in \{\su, \sd\}$.
The first sum runs over all nearest neighbor pairs, $\<m,n\>$, such that $|n-m|=1$ or $M-1$ at the periodic boundary.
Additionally, we consider the environment in the form of a pool of identical fermionic baths with the energy dispersion relation $\eps_k$, described by the bath Hamiltonian
\begin{align}
  H_B &= \sum_{m,s} H_{B,m,s}, &
  H_{B,m,s} &= \sumint_k \eps_k\, b\s_{m,s,k}\, b_{m,s,k},
\end{align}
with $b\s_{m,s,k}$ and  $b_{m,s,k}$ the fermionic creation and annihilation operators for electrons at bath $m \in \{1, ..., M\}$ with spin $s \in \{\su, \sd\}$ in mode $k$,
coupled separately, each bath to one site, by the tunneling amplitudes $\ga_{m,s,k}$
between the $k$--state in bath $m$ with spin $s$ and an electron at site $m$ with spin $s$,
described by the interaction Hamiltonian
\begin{align}
  H_I &= \sum_{m,s} H_{I,m,s}, &
  H_{I,m,s} &= \sumint_k \ga_{m,s,k}\, c\s_{m,s}\, b_{m,s,k} + \hc
\end{align}
The baths are 
at thermal equilibrium and satisfy $\<b\s_{m,s,k} b_{m',s',k'}\> = \d_{mm'} \d_{ss'} \d_{kk'}\, f(\eps_k)$ 
with the Fermi--Dirac distribution $f(E) = [1+\exp(\beta (E-\EF)]^{-1}$, the inverse temperature $\beta = 1/(k_B T)$ and the chemical potential $\EF$. In the following, we set the Boltzmann constant $k_B=1$ and the chemical potential $\EF = 0$, for convenience. 
The total Hamiltonian is then given by $H = H_S + H_B + H_I$.


The Hubbard system \eqref{Hubbard} itself has been extensively studied in the last decades and its properties are widely known 
\cite{lieb1989a, tasaki1998a, essler2005, qin2022a, arovas2022a, schadschneiderxxxx}. 

\subsection{System properties} \label{sec:Energy-Sectors}

\begin{figure}[ht]
  \begin{center}
    \includegraphics[width=\linewidth]{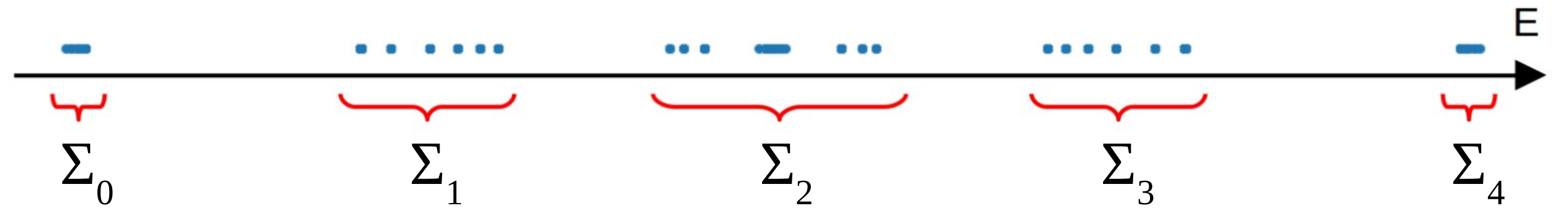}
  \end{center}
  \caption{Example of energy sectors $\Si_l$ for $M=4$ with $U \gg J$.
  We study relaxation from the excited, high--energy sectors, $\Si_l$ with $l>1$ towards the low--energy sector $\Si_0$.
  \label{fig:E-sectors}}
\end{figure}

First, we want to consider the properties of the system isolated from the environment ($\Ga = 0$). 
The choice $\eps = -U/2$ in \eqref{Hubbard} will bring some computational simplifications%
\footnote{The exact value of $\eps<0$ is not crucial but the symmetric choice  is more convenient for the study of the relaxation to the half filling sector with one electron per site, on average.},
namely the empty (holon) and doubly occupied (doublon) sites will have the same on--site energy $E_H = E_D = 0$ whereas the single electron occupations will have a negative energy $E_1 = -U/2$.
The system Hamiltonian \eqref{Hubbard} can be then split into $H_S = H_J + H_U + H_0$
with
\begin{align} \label{H_SJU}
  H_J &= J \!\!\! \sum_{\<m,n\>, s} c\s_{m,s} c_{n,s}, &
  H_U &= \frac{U}{2} \NDH, & 
  H_0 &= - \frac{M U}{2},
\end{align}
representing the kinetic $H_J$, on--site $H_U$ (together $H_{JU} \equiv H_J + H_U$) and constant $H_0$ parts,
with the total number of doublons and holons
\begin{align} \label{NDH}
  \NDH &= \sum_m \left[ n_{m,\sd} n_{m,\su} + (1 - n_{m,\sd}) (1 - n_{m,\su}) \right].
\end{align}
The spectrum of $H_J$ is bounded by $4 J M/\pi$, $H_U$ is non--negative and $H_0$ can be ignored as it has no influence on the dynamics. 
The total number of doublons and holons $N_{DH}$ has eigenvalues $0, 1, 2, ..., M$. 
In consequence, by the Weyl theorem \cite{Weyl-Th}, for large enough $U \gg J$ (more precisely, if $U/J > 16 M / \pi$), the spectrum of $H_S$ splits into a series of $M+1$ disjoint intervals $\Si_l$, $l=0, 1, ..., M$ (sectors with $N_{DH} = l + \Ocal{J/U}$) of width not exceeding $8 J M/\pi$, cf. Fig.~\ref{fig:E-sectors}.

When connected to fermionc baths it is expected that the state of the system will undergo a relaxation process in which, by exchanging particles and energy with the environment, it will settle down to some universal final state%
\footnote{One or a set of final states in the case of energetic degeneracy or trapping, see below for the discussion.}. 
It is anticipated that the cold baths will absorb most of the system's energy and thus force the system into its low--energy sector as far as such processes are possible.
For strong on--site repulsion $U>0$ (with hierarchy $T, \Ga \ll J \ll U$),
it should {induce} relaxation towards the half--filling sector, $\Si_0$, with on average one electron per site, as empty (holons) and double occupied sites (doublons) have higher on--site energies.
The final phase of relaxation should then take place only within the low energetic sector ($\Si_0$) where {the system} can be approximately described by the Heisenberg Hamiltonian coupled effectively to the environment. 
Since the final relaxation will be able to proceed only via energetically forbidden states (lying outside of the low--energy sector $\Si_0$), its rate should be much slower than that of the initial pre--relaxation phase and scale at least as $J/U \ll 1$.
We will show that different approaches and approximations may lead to different decay mechanisms and final states
what presents a warning for a too easy justification of a chosen approximation scheme.

\subsection{Lindblad master equations} \label{sec:Lindblad}

Together with the baths, the evolution is quite complex and takes place in an infinite dimensional Fock--space.
Since we are finally interested in the system's dynamics only we will eliminate the baths' degrees of freedom and derive a master equation for the Hubbard--system only. 
In the first step, we derive the Redfield equation which acts as a first approximation 
but has some mathematical problems \cite{Spohn-PositivityPreserving, B07-Dimer, Hatrmann+Strunz-RedfieldNonpositivity}. 
In the next step, we go over to Lindblad equations which should resolve them.
Here, we only highlight the main steps, while the details of the derivation are provided in App. \ref{app:Lindblad}.

The full system begins its evolution at time $t=t_0$ in a product state described by the density matrix
$ \rho(t_0) = \rho_S(t_0) \otimes \rho_B(t_0) $,
where $\rho_S$ refers to the Hubbard system while $\rho_B$ refers to the baths at thermal equilibrium. 
The evolution is governed by the von Neumann equation, 
$ \p_t\, \rho(t) = -i\, [H, \rho(t)] $ 
where we introduced the convention $\hbar = 1$, fixing the energy unit.

By applying the Born and the (first and second) Markov approximations to the bath and system evolution and
tracing out the baths' degrees of freedom in the von Neumann equation, expanded to the lowest non--vanishing order in the system--bath couplings, $\Ocal{|\ga_{m,s,k}|^2}$,
we arrive at a master equation known as the (second) Redfield equation \cite{Redfield, redfield1965a}
\begin{equation} \label{Redfield}
   \p_t\, \rho(t) = -i\, [H_S, \rho(t)] + \L\, \rho(t). 
\end{equation}
The superoperator
$\L$ can, in general, be not posi\-tivity preserving (short \textit{not positive}) which may lead to negative or unbounded probabilities \cite{Spohn-PositivityPreserving, B07-Dimer, Hatrmann+Strunz-RedfieldNonpositivity, LL-MTh, EK-PhD}. 
Therefore, it is usually corrected to the Lindblad form \cite{Manzano-LindbladEq, GSchaller-Review-NQSys}
which restores the positivity. 

\subsubsection{Global secular approach}

The most popular method is the \textit{secular} 
approximation \cite{Breuer, GSchaller-Review-NQSys} in which
$\L$ can be compactly written, cf. \cite{B07-Dimer}, as 
\begin{equation} \label{LL-sec-Lind}
  \L^\text{sec}\, \rho =  \sum_{\substack{m,s,\al \\ \D E}} \Ga_{m,s}\, \mathring \L[L_{m,s}^{\al,\text{sec}}(\D E)]\, \rho
\end{equation}
with $\alpha = \pm$ and the sum running over all energy differences $\D E = E_i - E_j$ in the spectrum of $H_S$
and using the Lindblad superoperator form
$  \mathring \L[L]\, \rho \equiv L\, \rho\, L\s - \frac{1}{2} \{ L\s L, \rho \} $
with a set of secular Lindblad jump operators 
\begin{multline} \label{Lind-glob-sec}
  L^{\pm,\text{sec}}_{m,s}(\D E) =  \\ \sum_{i,j} \d_{\pm \D E, E_i-E_j} \sqrt{f_+(E_i-E_j)}\, \ketbra{\chi_i}{\chi_i} c^\pm_{m,s} \ketbra{\chi_j}{\chi_j}
\end{multline}
where $f_\pm(\om) \equiv f(\pm \om)$, $c_{m,s}^+ \equiv c\s_{m,s}$ and $c_{m,s}^- \equiv c_{m,s}$.
The double sum runs over all eigenstates $\ket{\chi_l}$ of the system Hamiltonian $H_S$ with eigenenergies $E_l$.
The Lindblad superoperator $\L$ is positivity preserving.
The prefactors $\Ga_{m,s}$ in \eqref{LL-sec-Lind} are obtained from the coupling parameters $\ga_{m,s,k}$ 
via $\Ga_{m,s}(\omega) = 2 \pi {\int \hspace{-8.5pt} \Sigma}_k |\ga_{m,s,k}|^2 \d(\om-\eps_k)$ 
and simplified, for dense {and homogeneous} spectra of the baths, 
by the \textit{wide--band limit},
$\Ga_{m,s}(\omega) \ra \Ga_{m,s}$.
We will also, for better readability, assume below spin and site--independent tunneling rates, 
i.e. $\Ga_{m,s} \ra \Ga$.
The generalization to $\Ga_{m,s}$ is straightforward.

\subsubsection{Global coherent approach} \label{sec:L-glob-coh}

Because the secular approximation is quite a drastic modification, losing almost all information about system coherences,
we developed in \cite{B07-Dimer} the \textit{coherent} approach as a much less invasive method of restoring the positivity.
It relies on the minimally invasive refinement of some off--diagonal matrix elements appearing in the Redfield superoperator $\L$, cf. \eqref{Redfield-matrix} in App. \ref{app:Lindblad}, 
including the arithmetic means of the Fermi--Dirac functions at different energies,
which in the secular approximation are set to zero, 
and replaces them by their geometric means, cf. also 
\cite{Breuer-CohLind, Kirsanskas+Wacker-CohLindPhenom, Davidovic2020} and \cite{Nathan-CohLindDerivation} for  the derivation.

In this way, 
the Liouville superoperator $\L$ from \eqref{Redfield} can be again written in the form \eqref{LL-sec-Lind} with now the \textit{coherent} Lindblad jump operators
\begin{equation} \label{Lind-glob-coh}
  L^{\pm,\text{coh}}_{m,s} = \sum_{i,j} \sqrt{f_+(E_i-E_j)}\, \ketbra{\chi_i}{\chi_i} c^\pm_{m,s} \ketbra{\chi_j}{\chi_j}.
\end{equation}
Their name is motivated by the fact that they are equal to the coherent sums of the secular jump operators, $L^{\pm,\text{coh}}_{m,s} = \sum_{\D E} L^{\pm,\text{sec}}_{m,s}(\D E)$, over the spectrum of energy differences.
The \textit{global coherent} Lindblad operators {thus} allow, 
through coherent sums {in}  \eqref{Lind-glob-coh}, for arbitrarily localized states whereas in the \textit{global secular} Lindblad operators all electrons tunnel from the baths non--locally into the system. 
The latter is {one of} the largest disadvantages of the secular approximation when applied to the description of the dynamics and usually offers a reasonable approximation only at late times.
However, in the case of a trimer, we will point to significant discrepancies between the secular and coherent schemes, {in particular} in the context of the final states {and stationary currents \cite{Breuer-CohLind}}. 

\subsubsection{Local approach} \label{sec:L-loc}

In the case when the inter--site tunneling $J$ is smaller than the bath--site coupling $\Ga$, a further approximation can be carried out, namely the terms containing $\ga_{m,s,k} J$ and $J^2$ and {their higher powers} can be neglected, by which all Lindblad operators simplify to operators acting only locally on single sites.
In \cite{B07-Tetramer} we have given a well--founded derivation of the local Lindblad operators and discussed their application. 
Formally, the local Lindblad operators can be obtained by setting $J=0$ in the global Lindblad operators \eqref{Lind-glob-coh} ({or} in the secular version \eqref{Lind-glob-sec}, before applying the deltas $\d_{\pm \D E, E_i-E_j}$ separating the terms with different energy differences)
{thus leading to
two versions of the operators, 
{local} secular,
\begin{align} \label{Lind-loc-sec}
  \ell^\pm_{m,s,1} &= \sqrt{f_\pm(\eps)}\,c^\pm_{m,s}\,(1-n_{m,\bar s}), \nonumber \\
  \ell^\pm_{m,s,2} &= \sqrt{f_\pm(\eps + U)}\,c^\pm_{m,s}\,n_{m,\bar s},
\end{align} 
and {local} coherent,  
\begin{align} \label{Lind-loc-coh}
  \ell^\pm_{m,s} &= \ell^\pm_{m,s,1} + \ell^\pm_{m,s,2},
\end{align}
for the creation (+) or annihilation (-) of the first and second electron at the site $m$ with spin $s$ ($\bar s$ stands for the opposite spin to $s$), introduced in App. \ref{app:Lindblad}.
} 

Among others, the local Lindblad operators preserve the total spin direction $\vec S$ 
and lead to a non--Gibbs steady state \cite{hofer2017a, B07-Dimer, B07-Tetramer, GSchaller-Review-NQSys}. 
In \cite{B07-Dimer} and \cite{B07-Tetramer} we have discussed these properties in more detail. 

\section{Early relaxation: \text{towards the low--energy sector}} \label{sec:Early-Relax}

For cold baths at $T \approx 0$ the system is expected to relax towards low--energy states: the ground state and, possibly, a set of trapped states if further transitions to the ground state get effectively blocked%
\footnote{Thermal equilibration is guaranteed for {global} secular Lindblads {only} \cite{Davies-MarkovianMEs1, Kossakowski-Equilibration}.
All Lindblad superoperators act as contracting {maps}, hence bring the state continuously closer to the final state, but the set of final states may depend on particular Lindblads.
}.
The latter can happen because of to conserved quantities (often existing at exactly $T=0$) or when the transition to the lower lying states is only possible by tunneling via a higher excited state and thus energetically forbidden%
\footnote{The latter happens at the lowest order coupling with the environment where the Lindblad operators block energetically forbidden transitions completely. Taking into account higher orders in $\Ga$, which is not the objective of this work, might open additional decay channels.}. 

The relaxation process depends on the type of the Lindblad operators, local or global, and by this, on the regime of the coupling constants $\Ga$ and $J$ {\cite{Global-Local-Sec-2HarmOsc, B07-Dimer, B07-Tetramer, GSchaller-Review-NQSys}}.
Here, we will utilize the adjoint Lindblad master equation {\cite{Breuer}} for operators
\begin{equation} \label{Lind-A} 
\begin{split}
  \p_t\, {A(t)} &= i\, [H_S, {A(t)}] \\
  &+ \Ga \sum_\eta \left[ L\s_\eta\, {A(t)}\, L_\eta
  - \frac{1}{2} \left\{ L\s_\eta L_\eta, {A(t)} \right\}\right]
\end{split}
\end{equation}
(written with the multi--index $\eta$ comprising $(m, s, \pm)$ for simplicity, cf. \eqref{Lind-glob-sec}, \eqref{Lind-glob-coh}, \eqref{Lind-loc-sec} or \eqref{Lind-loc-coh}),
from which we obtain the time--dependent expectation value in the state $\rho$ by {noting}
\begin{equation}
  A(t) \equiv {\Tr[A(t) \rho(0)] = \Tr[A \rho(t)] 
  \equiv \ev{A}_{\rho(t)}}. 
\end{equation}
With this method, we can better understand the general dynamics of the observable ${A(t)}$ without referring to particularly chosen states $\rho$. 
The above equation takes an especially simple form for operators ${A(t)}$ which commute with the Hamiltonian $H_S$
{since} their dynamics is driven solely by the Lindblad jump operators. 
(Below, in this section, we skip the time index for simplicity but reinstate it when needed.)

\subsection{Relaxation with local Lindblads}

At $T=0$, for the total particle number $N$, the total spin $\v S$ (defined below, in \eqref{SpinOp}), spin squared $S^2$ and the energy $H_{JU}$, we can find the dynamical equations
\begin{align} \label{Nt-local}
  \p_t N(t) &= -2 \Ga (N(t) - M), \\ 
  \p_t S_i(t) &= 0, \\ \p_t S^2(t) &= \frac{2 \Ga}{U} H_U(t) = \Ga N_{DH}(t), \\
  \p_t H_{JU}(t) &= - 2 \Ga H_{JU}(t). \label{Ht-local}
\end{align}
{following from \eqref{Lind-A} with local Lindblad operators.} 
In direct consequence, $N(t) \ra M$ and $H_{JU}(t) \ra 0$ decay exponentially fast with the exponent $2 \Ga$, {cf. Fig. \ref{fig:N} and \ref{fig:H} (top)}. 
In contrast, $S^2(t) \ra S^2_\text{max} = \frac{M(M+2)}{2}$, $N_{DH}(t) \sim H_U(t) \ra 0$, $H_J(t) \ra 0$ decay at slower rates because their dynamics is more complicated, {cf. Fig. \ref{fig:N} and \ref{fig:H} (bottom)}.
Hence, {the steady states being} the final states of evolution have zero energy, ${\ev{H_{JU}}} = 0$, maximal spin, ${\ev{S^2}} = S^2_\text{max}$, and no doublons and no holons, ${\ev{N_{DH}}} = 0$, i.e. are exactly at half filling, also locally at each site.

The only states with exactly one particle per site and maximal spin are those with all spins pointing in the same direction (with $M+1$ possible {eigenvalues} of $S^z$ from $-M/2$ to $+M/2$). 
In these states, the hopping is blocked by the Pauli exclusion principle and therefore also ${\ev{H_J}} = 0$.
At local half--filling, ${\ev{H_U}} = 0$ and hence ${\ev{H_{JU}}} = 0$, too.

It is important to note that ${\ev{H_{JU}}} = 0$ is not the global energy minimum, since the energy ${\ev{H_{JU}}}$ can become negative. Driving the system into that particular energy is a property of the local Lindblad operators. 
It is known that also for finite temperatures the local Lindblads do not lead to the Gibbs state \cite{hofer2017a, B07-Dimer}. 

\subsection{Relaxation with global Lindblads}

\begin{figure}[t]
  \begin{center}
    \includegraphics[height=5.5cm]{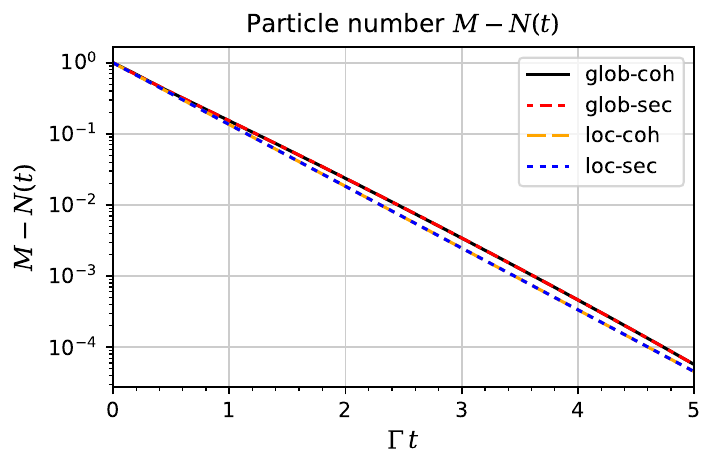} 
    \includegraphics[height=5.5cm]{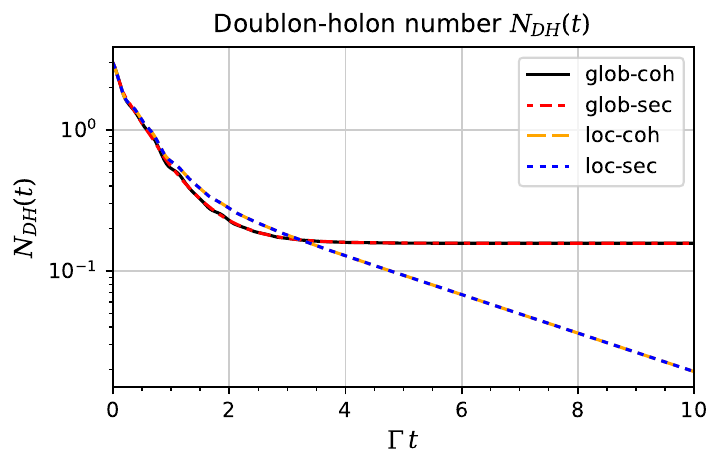} 
  \end{center}
  \caption{Top: Total particle number. Plotted is the difference $M-N(t)$. Different Lindblad operators compared here show no significant differences.
  Bottom: Total number of doublons and holons. The decay has a much slower tail and is different for different Lindblad operators: it tends to zero for local and to a non--zero value for global.
  Both plots for $M=3$, $J = 2\, \Ga, U = 10\, \Ga, T=0$ and the initial state populated with two electrons in one site, $\ket{\psi_0} = \ket{\sd\su, 0, 0}$. 
  \label{fig:N}}
\end{figure}

For observables evaluated with the global Lindblad equation, the situation is more complex 
since the {global} Lindblad operators have a more complicated form and the multiple sums in \eqref{Lind-A} cannot be evaluated as easily as in the local case (each $L_\al$ involves a further double sum). 
{
However, observing that the mismatch between the local (w.r.t. $H_U$) and global (w.r.t. $H_S$) eigenstates, defining the Lindblad operators, is small of the order of $J/U$,
we are able to show that the early global relaxation proceeds similarly to the local one 
(cf. App. \ref{app:Nt-Lind} for more details).
The total particle number satisfies 
}
\begin{equation}
  N(t) \approx M + [ N(0) - M ]\, e^{-2\Ga'\, t}
\end{equation}  
with a slightly modified exponent $\Ga' = \Ga + \O{J/U}$, cf. Fig. \ref{fig:N} (top).
Thus, $N(t)$ is very close to the analytic solution obtained for the local Lindblads \eqref{Nt-local}.
Also the numbers of doublons and holons $\NDH$ stay close to each other for various Lindblads in the early relaxation stage (but eventually differ), cf. Fig. \ref{fig:N} (bottom).

\begin{figure}[t]
  \begin{center}
    \includegraphics[height=5.5cm]{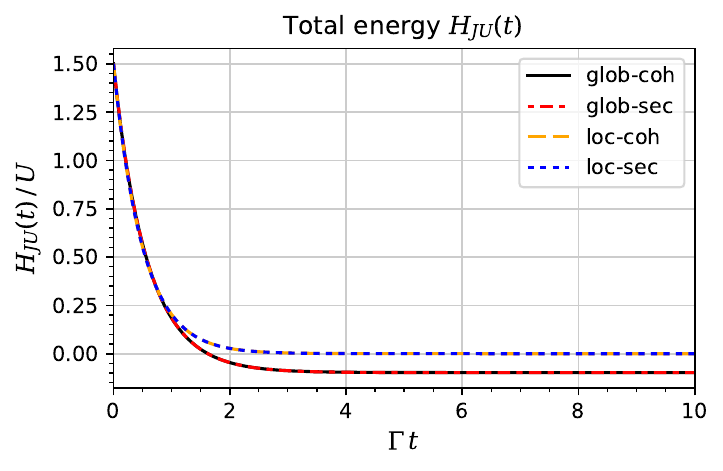} 
    \includegraphics[height=5.5cm]{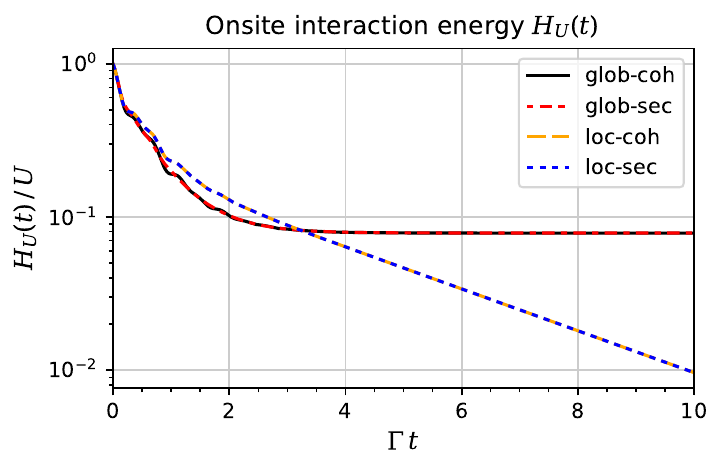} 
  \end{center}
  \caption{Top: Total system energy. Plotted is $H_{JU}(t) = H_S(t) - H_0$. Different Lindblad operators compared here show no significant differences until the total energy relaxes to the scale of $\sim M J = 6\, \Ga$ ($H_{JU}/U \sim 0.6$). 
  Bottom: Total interaction energy $H_U(t) \sim \NDH(t)$. The decay has a much slower tail and depends on the approximation model: it tends to zero for local Lindblads and to a non--zero value for global Lindblads.
  (Parameters and the initial state as in Fig.~\ref{fig:N}.)
  \label{fig:H}}
\end{figure}

{The total energy of the system, for $T=0$, satisfies}
\begin{equation}
  H_{JU}(t) = H_{JU}(0)\, e^{-2\,\Ga t} + \Ocal{2^M J},
\end{equation}
which means that the early relaxation governed by the global Lindblads proceeds closely to that governed by the local Lindblads, as long as the energy is in the higher energy sectors, $\Si_1, \Si_2, ...$ with ${H_S(t)} \gtrsim U \gg 2^M J$.
{Once} it reaches the lowest band $\Si_0$ at the order of $J$ it starts to significantly differ (cf. Fig. \ref{fig:H}). 
We have no sufficient tools to describe the difference more precisely here but will do this below within the Heisenberg approximation valid in the lowest sector $\Si_0$.

In conclusion, we did not observe any significant differences in the discussed variables between the coherent and secular approximations (at $T=0$) for either the local or the global Lindblads
during the early state of relaxation. 

\section{Properties of the low--energy sector} \label{sec:Low-Energy}

\subsection{The Heisenberg Hamiltonian}

\begin{figure}[t]
  \begin{center}
    \includegraphics[width=1.0\linewidth]{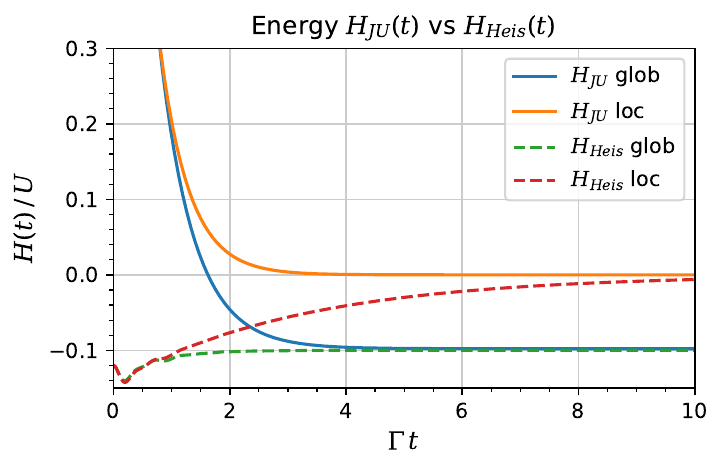}
  \end{center}
  \caption{The energy $H_{JU}(t)$ compared with the Heisenberg energy $H_\Heis(t)$ for local and global Lindblad operators (secular and coherent not distinguishable). 
  They converge pairwise together when the number of doublons and holons $\NDH$ becomes small (cf. Fig. \ref{fig:H}).
  (Parameters and the initial state as in Fig. \ref{fig:N}.)
  \label{fig:Heis}}
\end{figure}

Here, we consider the properties of the system in the lowest energy sector, again isolated from the environment. 
The lowest energy sector $\Si_0$, with energies $E \ll U$ consists of states 
with {only} small contributions of doublons and holons,
suppressed by $J/U$.
These states can be interpreted as ``dressed'' eigenstates $\ket{\psi_k}$ of the Hubbard Hamiltonian $H_S$
and obtained by small unitary rotations of ``bare'' states $\ket{\t \psi_k}$  having all a fixed number of doublons and holons.
For this, we apply the \textit{modified UDL decomposition}, proposed by us, cf. App \ref{app:UDL}, which replaces the standard Schrieffer–Wolff transformation \cite{Schrieffer-Wolff} and transforms the states and the Hubbard Hamiltonian according to
\begin{equation} \label{UDL}
  \ket{\t \psi_k} = W\s \ket{\psi_k}, \qquad \t H_S = W\s H_S W 
\end{equation}
with some unitary $W = \1 + \Ocal{J/U}$. 
In the lowest energy sector and for $J \ll U$ the Hubbard Hamiltonian can be approximated this way by the Heisenberg Hamiltonian \cite{cleveland1976a},
{$ \left. (H_{JU} + H_0) \right|_{\Si_0}  = \left. H_S \right|_{\Si_0} \approx \left. \t H_S\right|_{\Si_0} \equiv H_\Heis + H_0$,}
\begin{equation} \label{Heis}
  H_\Heis = \frac{4 J^2}{U} \sum_{\<m,n\>} \left( \v S_m \cdot \v S_n - \frac{1}{4} \right) 
\end{equation}
with the exception for $M=2$ where the coefficient in front%
\footnote{Hopping over the boundary to the same site in the periodic $M=2$ system doubles the number of second order processes in $H_J$ contributing to the effective Heisenberg Hamiltonian.}
is $8$ instead of $4$ and
with the local spin operators 
\begin{equation} \label{SpinOp}
  \v S_m = \frac{1}{2} \sum_{s,s'} c\s_{m,s}\, \boldsymbol{\si}_{s,s'}\, c_{m,s'}
\end{equation}
where $\boldsymbol{\si}$ represents the Pauli matrices $\si^1, \si^2, \si^3$.
Since the nearest neighbor spin order operator $O_1$, appearing in the Heisenberg Hamiltonian, is bounded from below and from above
\begin{equation}
  O_1 \equiv \sum_{\<m,n\>}  \v S_m \cdot \v S_n
  \quad 
  \text{with}
  \quad 
  \frac{-3M}{4} \leq O_1 \leq \frac{M}{4},
\end{equation}
it immediately follows%
\footnote{There is no contradiction in the fact that the spectrum of $\t H_S$ appears narrower than that of $H_S$ in the lowest energy sector since the latter was only estimated based on separate estimates of $H_J$ and $H_U$ (cf. below Eq. \eqref{NDH}), without taking into account that some combinations of both may not be possible.} 
that $-4J^2M/U \leq H_\Heis \leq 0$. 
The validity of the approximation \eqref{Heis} is demonstrated in Fig. \ref{fig:Heis}.

\subsection{Spin waves}

In the transformed eigenbasis $\ket{\t \psi_k}$ of $H_\Heis$,
the lowest energy sector $\Si_0$ consists only of states with 
{local half filling, $n_m = 1$, i.e. one electron per site and no doublons and holons, ${{N_{DH}}} = 0$}.
The only local degree of freedom is the spin and therefore the sector consists of $2^M$ states. 
It is convenient to categorize them by the quantum numbers corresponding to the operators 
$S^2, S^z, O_1$ and $I_S$ (defined below) which all commute 
with the Heisenberg Hamiltonian $H_\Heis$.

{
The states with  maximal spin order $\ev{O_1} = \frac{M}{4}$, where all spins point in the same direction, are eigenstates of the Heisenberg Hamiltonian with 
maximal energy, ${{H_\Heis}} = 0$. 
Keeping in mind the overall $SU(2)$ spin symmetry of the system, we choose an arbitrary direction, e.g. the $z$--axis, and study spin flips along it. 
For $\ket{Z^+} = \ket{\su, ..., \su}$ and $\ket{Z^-} = \ket{\sd, ..., \sd}$, 
where $\su, \sd$ denote spin ``up'' and ``down'' along the $z$--axis
we have $\ev{S^z}_{Z^\pm} = \pm \frac{M}{2}$
and the hopping is blocked by the Pauli principle, hence they are non--dynamic.
}

For the investigation of {further spin states, it is convenient} to introduce local spin--flip operators at site $m$
\begin{equation} \label{spin-ladder-ops}
  S^+_m = c\s_{m,\su} c_{m,\sd}, \qquad S^-_m = c\s_{m,\sd} c_{m,\su}.
\end{equation}
{An interesting spin dynamics can be observed already in the sector with one flipped spin}, $\ket{S^\pm_m} \equiv S^\mp_m \ket{Z^\pm}$ 
{(called in the literature \textit{magnons} \cite{Kittel-Book}), in which} a distinguished role is played by the basis
of spin waves 
\begin{equation} \label{spin-waves}
\begin{split}
    \ket{\t S^\mp_k} &= \t S^\pm_k \ket{Z^\mp} = \frac{1}{\sqrt{M}} \sum_m e^{i k m} S^\pm_m \ket{Z^\mp} 
\end{split}
\end{equation}
{with the Fourier transforms $\t S^\pm_k$}
for the quantized spin wavenumbers $k = 2 \pi s / M$, $s \in \{0, 1, ..., M-1\}$.

{
The spin order and Heisenberg Hamiltonian can be diagonalized in the spin wave basis
$\ket{\t S_k^\pm}$ with the eigenvalues}
\begin{align} 
  \ev{O_1}_k &= \cos(k) + M/4 - 1, \label{O_k} \\
  \ev{H_\Heis}_k &= \frac{4J^2}{U} \left( \cos(k) - 1 \right) \leq 0 \label{Heis_k}
\end{align}
(cf. App. \ref{app:Spin-waves} for the discussion and calculations).

\subsection{Particle and spin currents}

{In the original system \eqref{H_SJU}, if} $U=0$, the hopping of particles is described only by $H_J$ 
and so the total number of particles $N$ is conserved, $[ H_J, N] = 0$, 
as well as the corresponding particle current
\begin{equation} \label{J_N}
  J_N = \sum_{m, s} i(c\s_{m+1,s} c_{m,s} - c\s_{m,s} c_{m+1,s})
\end{equation}
with $ [H_J, J_N] = 0 $.
Moreover, each component of the total spin $\v S$ is conserved, $[ H_J, \v S] = 0$, 
and there exists a corresponding conserved spin polarized current
\begin{equation} \label{J_S}
  \v J_S = \frac{i}{2} \sum_{m,s,s'} \left(c\s_{m+1,s} \boldsymbol{\si}_{s,s'} c_{m,s'} - c\s_{m,s} \boldsymbol{\si}_{s,s'} c_{m+1,s'} \right)
\end{equation}
which satisfies $ [H_J, \v J_S] = 0$, too.
{In the low energy sector of the full system,}
the UDL--trans\-formation \eqref{UDL} maps the original currents $J_N$ and $\v J_S$ onto {$\t J_N$ and $\t{\v J}_S$} valid in the low--energy sector.
The transformed particle current $\t J_N = 0$ vanishes because there is no particle motion in the Heisenberg sector
{
while the total spin and the spin current 
satisfy the conservation laws
\begin{align}
  [H_\Heis, \v S] &= 0, & [H_\Heis, \t{\v J}_S] &= 0. 
\end{align}
The latter is, however, only exact for $M=2, 3$ while only approximate for $M\geq 4$.
The spin current $\t{\v J}_S$ can be rewritten as a dimensionless operator
\begin{equation}
  \v I_S \equiv - U \t {\v J}_{S} = 2 \sum_m \v S_m \times \v S_{m+1}
\end{equation}
and its $z$--component can be expressed in terms of the spin--flip operators $S^\pm_m$  
\begin{equation} \label{I_S}
  I_S^z = \sum_m i(S^+_{m+1} S^-_m - S^+_m S^-_{m+1})
\end{equation}
(cf. App. \ref{app:Spin-current} for derivations and discussion).
}

{Thus}, the single spin waves $\ket{\t S^\pm_k}$ are eigenstates of the spin--$z$ $S^z$, spin order $O_1$, and the spin current operator $I_S^z$ with the eigenvalues of the latter
\begin{equation} \label{I_S-eig}
  \ev{I_S^z}_k = 2 \sin(k).
\end{equation}

\subsection{Magnons}

The UDL--transformation \eqref{UDL} is unitary 
and maps the energy eigenstates of $H_S$ onto the energy eigenstates of $H_\Heis$. This mapping can be now inverted to map back the Heisenberg spin wave states, including the ground state(s), onto the  eigenstates, including the ground state(s), of the original Hubbard system. 
In the latter context, we will call the eigenmodes of the interacting Hubbard Hamiltonian carrying the spin quasimomentum and the spin current \textit{magnons} in order to differentiate them from the effective  Heisenberg \textit{pure spin wave modes}.
Both, the (single) spin waves and the magnons form orthonormal bases.

By the inverse (UDL) mapping, from the Heisenberg to the Hubbard model, the global ground state(s) of the Hubbard Hamiltonian $\ket{\psi_*}$ should consist of magnons
corresponding to the spin waves $\ket{S_*}$ with non--vanishing wavenumbers $k \neq 0$ with negative Heisenberg energy, ${\ev{H_\Heis}_k} \sim \cos(k) - 1 < 0$.
Since the UDL--transformation preserves the scalar products and thus also the expectation values of operators, the energy of the magnonic ground state(s) $\ket{\psi_*}$ reads
\begin{align} \label{E-magnons}
  E_* &= \ev{H_S}_{\psi_*} 
  = H_0 +  \ev{H_\Heis}_{S_*} 
  + \Ocal{\frac{J^3}{U^2}}.
\end{align}
Their spin current expectation values
\begin{equation} \label{IS-magnons}
  \ev{I_S^z}_{\psi_*} = |\braket{S_* | \psi_*}|^2 \ev{I_S^z}_{S_*}
\end{equation}
as well as the order
\begin{equation} \label{O1-magnons}
  \ev{O_1}_{\psi_*} = |\braket{S_* | \psi_*}|^2 \ev{O_1}_{S_*}
\end{equation}
can be obtained by the projections $\braket{\psi_*|S_*}$
since the doublon--holon mixtures included in the difference $\ket{\psi_*} - \ket{S_*}$ do not contribute to these observables.

Since the condition $\cos(k) - 1<0$ can be satisfied only by $k\neq 0$, the ground state needs to correspond to magnonic modes $\ket{\psi_k}$ carrying non--vanishing spin current proportional to $\sin(k)$, cf.~\eqref{I_S-eig}.
Since the global Lindblad operators for $T \approx 0$ push the system towards its ground state we expect the spin current carrying magnons, $\ket{\psi_k}$, to play the role of global attractors of evolution.
We expect persistent (or long living) spin currents, despite the decoherence owing to the interaction of the system with the environment.

\section{Late relaxation: \text{inside the low--energy sector}} \label{sec:Late-Relax}

The relaxation process within the low--energy sector depends strongly on the type of the Lindblad operators, local or global, and by this, on the coupling regime.
In principle, the Lindblad operators $L_\alpha$ derived for the original system should be unitarily ``rotated'' to $\t L_\alpha = W\s L_\alpha W$, cf.~\eqref{UDL}, from the eigenstate basis of the Hubbard Hamiltonian $H_S$ to the eigenstate basis of the Heisenberg Hamiltonian $H_\Heis$. 
{Such adapted Lindblads should describe the relaxation processes in the low--energy sector described by the Heisenberg Hamiltonian.} 
However, when restricted to the low--energy subspace, processes in which one electron is added or removed to the system fall out because of the projection procedure onto $\Si_0$ and thus we would miss the relevant relaxation mechanisms%
\footnote{The derivation of the Redfield or Lindblad equation, cf. App. \ref{app:Lindblad}, relies on quadratic terms in the system--bath coupling Hamiltonian, $H_C^2$, and the proper relaxation channels are restored when the projection onto the low--energy sector is performed first on $H_C^2$, not on $H_C$.}.
By including second order exchange processes with the baths and eliminating again the baths we can arrive at further relaxation paths, including spin--flips or spin--hopping processes.
However, this would require a quite involved derivation of the new set of ``second order'' Lindblad operators which we leave to further investigation. 
Instead, we use here the consequences following directly from the properties of the original Lindblad operators, switching between the properties of the original Hubbard and the effective Heisenberg system. 

\subsection{Relaxation with local Lindblads}

\begin{figure}[t]
  \begin{center}
    \includegraphics[height=5.5cm]{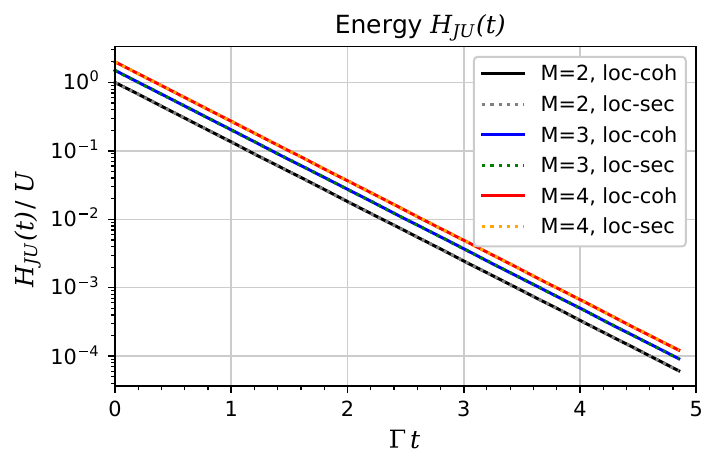} 
    \includegraphics[height=5.5cm]{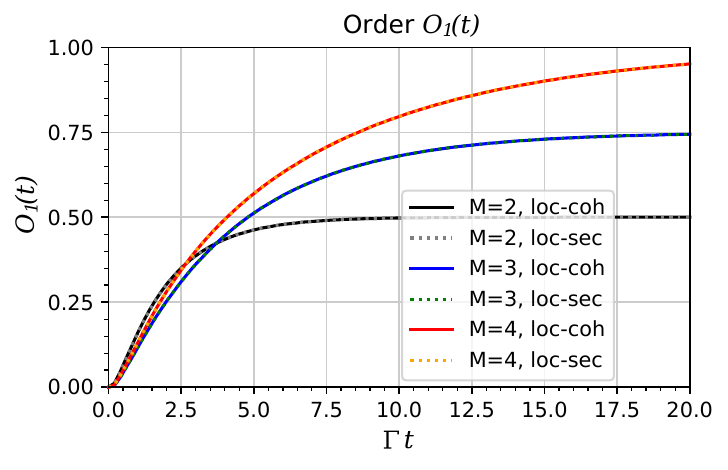} 
  \end{center}  
  \caption{In relaxation governed by the local Lindblads, the energy $H_{JU}(t)$ (top) decays exponentially with the rate $2\,\Ga$ towards zero.
  The spin order $O_1(t)$ (bottom) increases towards its maximum $M/4$ for systems of the size $M = 2, 3, 4$. 
  Both plots for $J = 2\, \Ga, U = 12\, \Ga, T=0$ and the initial states are empty, $\ket{\psi_0} = \ket{0, ..., 0}$.
  \label{fig:E+O-LateRelax-Loc}
  }  
\end{figure}
 
\begin{figure}[t]
  \begin{center}
    \includegraphics[height=5.5cm]{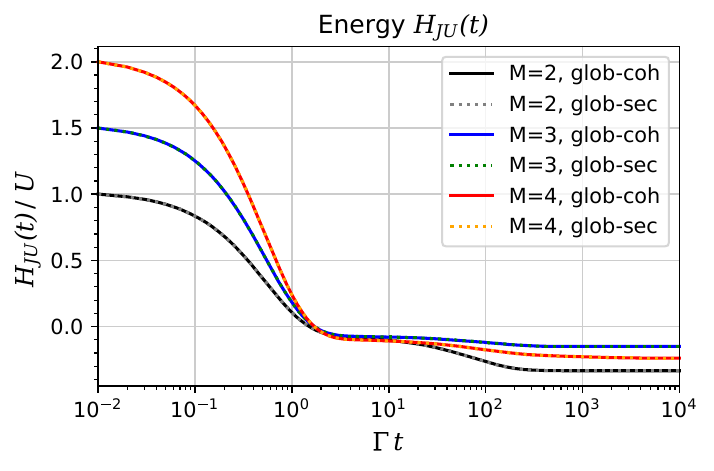} 
    \includegraphics[height=5.5cm]{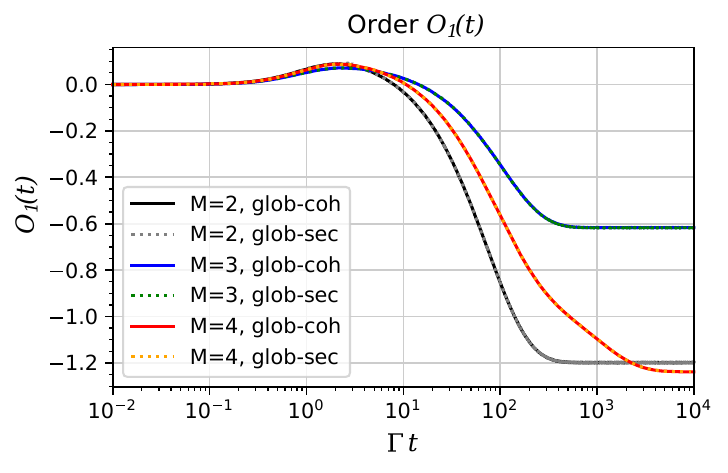} 
  \end{center}
  \caption{In relaxation governed by the global Lindblads the energy $H_{JU}(t)$ (top) decays initially exponentially with the rate $2\,\Ga$ and finally with the rate $\Ga \exp[-\D E/T]$  for $T>0$ where $\D E = U/2 + \Ocal{J}$ is the energy separation between the lowest energy sector $\Si_0$ and the first sector $\Si_1$. 
  For the Heisenberg model, the ground state energy is $\ev{H_\Heis}_\text{min} = -16J^2/U, -6 J^2/U, -12 J^2/U$ while
  the spin order $O_1(t)$ reaches its theoretical minimum at $\ev{O_1}_\text{min} = -\frac{3}{2}, -\frac{3}{4}, -2$ for systems of the size $M = 2, 3, 4$ sites, respectively.   
  Because of the finite temperature $T = 0.5\, \Ga$ and finite ratio of $J/U$, which are necessary for the observation of the late decay at finite times, the observed Hubbard values deviate from the Heisenberg minima, as given in \eqref{E-magnons} and \eqref{O1-magnons} and discussed below, in Sec.~\ref{sec:Trimer-Magnons}, and also get perturbed towards the Gibbs distribution.
  (Other parameters and the initial states as in Fig.~\ref{fig:E+O-LateRelax-Loc}.)
  The parameters are chosen such that $\D E \approx 2\,\Ga$ for $M=2, 3$ and the late decay time--scale is $\Ga\,t \sim \exp(\D E/T) \approx 54.6$.
  For $M=4$ the late decay proceeds in two steps, with $\D E \approx 2\,\Ga$, $\D E' \approx 3.6\,\Ga$, because up to two spins must be re--ordered to reach the ground state
  which gives a secondary time-scale $\Ga\,t' \sim \exp(\D E'/T) \approx 1340$.
  (The logarithmic scaling of time is {chosen} for better presentation {of different time scales}. The evolution begins at $t_0 = 0$.)
  \label{fig:E+O-LateRelax-Non}
  }
\end{figure}

As shown above, under the action of local Lindblad operators, the total spin in the system becomes maximal 
and the order ${O_1(t)}$ in the system increases to its maximum. 
Since we do not present a direct proof here, an alternative way, based on the energy in the low--energy sector, can be used instead. 
The Hubbard Hamiltonian at low energies is well approximated by the Heisenberg Hamiltonian \eqref{Heis} and the total energy ${H_{JU}(t)}$ tends to zero at late times, cf.~\eqref{Ht-local}, while the Heisenberg energy is non--positive, ${H_\Heis(t)} \leq 0$, as stated above. Therefore, since the two need to coincide at late times, 
cf. Fig. \ref{fig:Heis},
the Heisenberg energy must reach its maximum value of zero, 
cf. Fig. \ref{fig:E+O-LateRelax-Loc} (top),
\begin{equation}
  {H_\Heis(t)} =  {\frac{4 J^2}{U} \sum_m \underbrace{\left[ \ev{\v S_m \cdot \v S_{m+1}}_t - \frac{1}{4} \right]}_{\leq 0}} \ra 0
\end{equation}
which is only possible when
\begin{equation}
  {\ev{\v S_m \cdot \v S_{m+1}}}_t \ra \frac{1}{4} \qquad \text{for all $m$}
\end{equation}
because of the upper bound $\v S_m \v S_{m+1} \leq \frac{1}{4}$. 
This indicates that all neighboring pairs of spins must be aligned and the system evolves towards the maximal spin order%
\footnote{This should not be confused with the strict \textit{ferromagnetic order} for here no particular direction is distinguished and the state can be in a superposition or mixture of aligned spins, e.g. 
$(\ket{\su ... \su} + \ket{\sd ... \sd})/\sqrt{2}$. 
The situation could be changed by adding a small constant magnetic field $\v B$ in a fixed direction with the Zeeman splitting Hamiltonian $H_Z = \v B \cdot \v S$ breaking the rotational symmetry.}, 
cf. Fig. \ref{fig:E+O-LateRelax-Loc} (bottom),
\begin{equation}
  {O_1(t)} = {\sum_m  \ev{\v S_m \cdot \v S_{m+1}}_t} \quad \stackrel{\max}{\longrightarrow} \quad  \frac{M}{4}.
\end{equation}

In the local approximation discussed here, which is more appropriate in the regime of weaker inter--site coupling ($J\lesssim \Ga$) where the environment introduces stronger decoherence,
the magnons will still be long living states as their decay is suppressed by the Coulomb blockade ($U \gg J, \Ga, T$).

\subsection{Relaxation with global Lindblads}

Under the evolution governed by the global Lindblads, the system 
at small but finite temperatures, $T\approx 0$,
relaxes towards the Gibbs state which is close to the global ground state%
\footnote{At exactly $T=0$, it can get trapped in some other state.} 
and the total system energy $H_S$ decreases towards its global minimum  $E_\text{min}$.

The Heisenberg energy $H_\Heis$ as well as the order $O_1$ also attain their minimal values,
cf. Fig. \ref{fig:E+O-LateRelax-Non},
\begin{equation}
  {H_\Heis(t)} = {\frac{4 J^2}{U} \sum_m \underbrace{\left[ \ev{\v S_m \v S_{m+1}}_t - \frac{1}{4} \right]}_{\geq -1}} \ra \text{min} 
\end{equation}
since $\v S_n \v S_{n+1} \geq -\frac{3}{4}$.
A minimum with all pairs simultaneously satisfying ${\ev{\v S_m \v S_{m+1}}_t =} -\frac{3}{4}$ is not achievable for $M>2$ because of the lower limit on the total $S^2$. 
Below, we look closer at some important examples.

\subsubsection{$M=2$}

For $M=2$ the ground state of the Heisenberg Hamiltonian $H_\Heis$ \eqref{Heis} is the singlet $\ket{\psi} = (\ket{\su,\sd} - \ket{\sd,\su})/\sqrt{2}$ with the minimal spin ${{S}} = 0$, order ${{O_1}} = -3/2$ and the Heisenberg energy ${{H_\Heis}} = -16J^2/U$.
From the perspective of the above defined spin waves%
\footnote{Here, the spin ordered backgrounds $\ket{Z^\pm} = \ket{T_\pm}$ are the extremal triplet states while the middle triplet $\ket{T_0} = S^\pm_0 \ket{Z^\mp}$ is a spin wave with wavenumber $k_0 = 0$.}
it is a wave with non--zero spin wavenumber $k_1 =\pi$,
i.e. $\ket{\psi} = S^\pm_1 \ket{Z^\mp}$.
However, because of the extreme simplicity of the two--site system {and the periodicity condition} in which the right neighbor is at the same time the left neighbor, this mode can carry no directed spin current which would in that case be proportional to $\sin(k_1) = 0$.
For this reason, it is necessary to consider at least three sites. 

\subsubsection{$M=3$}

For $M=3$ all spin pairs are nearest neighbors and we can therefore write
\begin{align} \label{S2_M=3}
  S^2 &= \sum_m \v S_m^2 + \sum_{m\neq n} \v S_m \cdot \v S_n \nonumber \\
  &= \frac{3}{4}M + 2 \sum_{m} \v S_m \cdot \v S_{m+1} = \frac{9}{4} + 2\, O_1.
\end{align}
The spin order ${{O_1}}$ gets hence minimal in the sector with the smallest total spin ${{S}} = \frac{1}{2}$ and ${{S^2}} = \frac{3}{4}$
which gives the minimal possible value ${{O_1}} = - \frac{3}{4} $.
For $M=3$ we only deal with single spin waves relative to the fixed ferromagnetic background
and the order {\eqref{O_k}} of the spin wave $\ket{\t S_k}$ is 
\begin{equation}
  \ev{O_1}_k = \cos(k) - \frac{1}{4}.
\end{equation}
It corresponds to the above given minimum of $-\frac{3}{4}$ for both propagating spin waves with $k=\pm k_0 = \pm 2\pi/3$.
The Heisenberg energy {\eqref{Heis_k}} becomes then 
\begin{equation}
  \ev{H_\Heis}_{\pm k_0} = - \frac{6 J^2}{U}.
\end{equation}
The non--propagating spin wave with $k=0$ has higher order $\ev{O_1}_{k=0} = \frac{3}{4}$ and higher energy $\ev{H_\Heis}_{k=0} = 0$.
Therefore, within the Heisenberg approximation, we should expect a global relaxation towards the two spin wave modes carrying non--zero spin current{, as discussed below, in Sec.~\ref{sec:Trimer}}. 

\subsubsection{$M=4$}

{For $M=4$ the analysis is more involved since the singlet (${{S}}=0$) ground state is realized by applying two spin flips on the ferromagenetic background $\ket{Z^\pm}$
what can be realized by two spin waves on top of each other. 
Therefore, we moved the discussion to the App. \ref{app:M=4} and state here only the results. 
}
The ground state has minimal spin ${{S}}=0$ and minimal order $ \ev{O_1}_{\rm af+} = - 2 $
which corresponds to the Heisenberg energy \eqref{Heis}
\begin{equation}
  \ev{H_\Heis}_{\rm af+} = - \frac{12 J^2}{U}.
\end{equation}
Therefore, within the Heisenberg approximation, we should expect a global relaxation to the two spin wave modes carrying partly (because of the $k = \pm \frac{\pi}{2}$ modes) non--zero spin current $\ev{I^z_S}_k \sim \sin(k)$, cf. \eqref{I_S-eig}.
This question for {$M \ge 4$} 
requires further systematic investigation.

\subsection{Local vs. global late relaxation}

Summarizing the above results, 
the relaxation based on the local Lindblads tends to a final state of maximally ordered spins (${S(t)} \ra S_\text{max}$, ${O_1(t)} \ra O_{1,\text{max}}$) such that no further hopping can take place (${H_J(t)} \ra 0$) and the local energy is minimized (${H_U(t)} \ra 0$).
The relaxation based on global Lindblads ends up in any state from the lowest energy sector $\Si_0$ from where no further decay at $T=0$ is possible%
\footnote{
All states $\ket{\psi}$ in the lowest sector and their coherences are then decoherence free  
because of $\sum_\al L\s_\al L_\al \ket{\psi} = 0$ for all global coherent Lindblad operators $L_\al$.
Global secular Lindblads will also preserve the sector but eliminate the coherences between the energy eigenstates unless they are degenerate in energy.}. 
At small positive $T > 0$ a slow decay towards the global energy minimum should occur.

\section{Trimer ($M=3$)} \label{sec:Trimer}

\begin{figure}[t]
  \includegraphics[width=\linewidth]{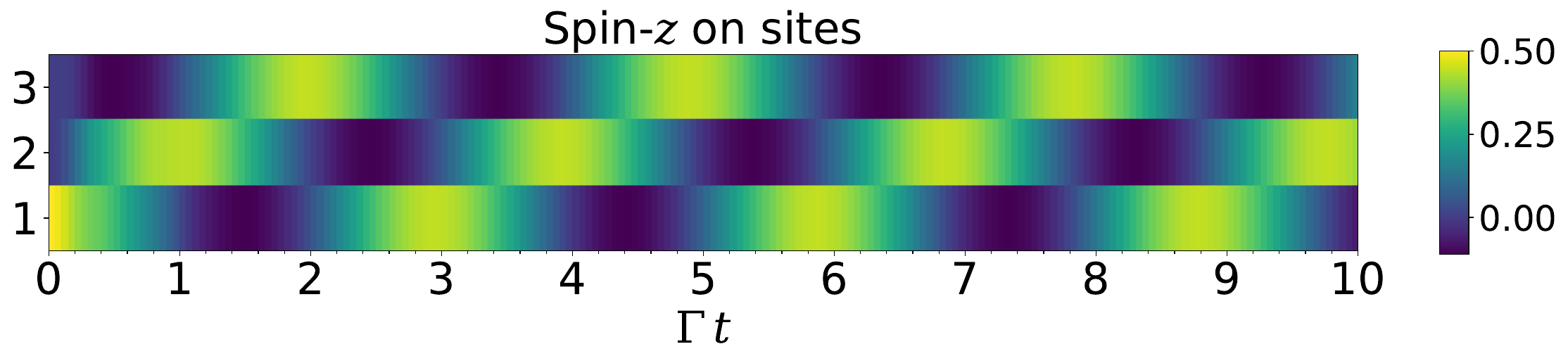}
  \caption{Spin--$z$ at three sites (1, 2, 3) as a function of time. Skew pattern (yellow) shows the propagation of a spin wave across the system.
  (Parameters and the initial state as in Fig.~\ref{fig:N}.)
  \label{fig:spincur_density}}
\end{figure}

Finally, we look closer at the trimer, with $M=3$ sites, which has the convenient property that only single spin waves are allowed in which one spin is flipped relative to the ferromagnetic ordered background. 
Moreover, for $M=3$ sites, the full Hubbard dynamics can be still treated numerically exactly in its full extent despite the dimension of the Hilbert space growing exponentially%
\footnote{The number of the degrees of freedom is \#sites $\times$ \#spins $= 2M$, hence the dimension of the wavefunction is $2^{2M}$, of the density matrix $2^{2M} \times 2^{2M} = 2^{4M}$ and of the Liouville operator $2^{4M} \times 2^{4M} = 2^{8M}$. For $M=3$ the latter is already $16.8\times 10^{6}$ and increases by the factor 256 with each additional site.}
with $M$.
The trimer is also the smallest system in which spin--waves, spin currents and relaxation to  ``dynamical''  states carrying persistent spin currents can be observed, cf. Fig. \ref{fig:spincur_density}.

\subsection{Spin waves}

{In the low energy sector governed by the Heisenberg Hamiltonian, } {the trimer has $2^M=8$ spin states: four (quadruplet) with the maximal spin $S=\frac{3}{2}$ and ferromagenetic order and four (two doublets) with $S=\frac{1}{2}$. 
In the spin wave basis \eqref{spin-waves}, the doublets 
correspond to spin waves $\ket{S^\ka_s}$ with spin momentum index $s=\pm 1$ and spin background index $\ka = \pm$
while the two of the quadruplet states with $|S^z|=\frac{1}{2}$ correspond to spin waves $\ket{S^\ka_0}$ with zero spin momentum, summarized in Tab. \ref{tab:spinwaves-alg}.
Altogether,}
the states $\ket{S^\ka_s}$ with $\ka = \pm$ and $s\in\{0,\pm 1\}$ 
{
carry spin--$z$ current \eqref{I_S-eig} $\ev{I^z_S}_{\ka,s} = \sqrt{3}\, \ka\, s$,
}
the spin order \eqref{O_k} $\ev{O_1}_{\ka,s} = -\frac{1}{4} - \frac{|s|}{2}$ and the Heisenberg energy \eqref{Heis_k} $\ev{H_\Heis}_{\ka,s} = - 2 |s| J^2 / U $, {as summarized in Tab. \ref{tab:spinwaves-kin}. 
The remaining two quadruplet states $\ket{Z^\ka}$ must have the same physical properties as the (quadruplet) spin wave states $\ket{S^\ka_0}$ because of the $SU(2)$ spin rotation symmetry.}
It follows immediately that the states $\ket{S^\ka_\pm}$ (with $s=\pm 1$) {have lower energy} than $\ket{S^\ka_0}$ (with $s=0$).
For more detailed discussion, cf. App. \ref{app:Trimer-Spin-states}.

\begin{table}[t] 
  \setlength{\tabcolsep}{12pt}
  \renewcommand{\arraystretch}{1.3}
  \begin{tabular}{|c|c|c|c|} \hline
    $S = \frac{3}{2}$ & $S = \frac{1}{2}$ & $S_z$ & \\ \hline
    $\ket{Z^+}$ &  & $+\frac{3}{2}$ & ferromagnetic  \\ \hline
    $\ket{S^+_0}$ & $\ket{S^+_\pm}$ & $ +\frac{1}{2}$ & spin waves \\ \hline
    $\ket{S^-_0}$ & $\ket{S^-_\pm}$ & $ -\frac{1}{2}$ & spin waves  \\  \hline
    $\ket{Z^-}$ &  & $ -\frac{3}{2}$ & ferromagnetic  \\  \hline
  \end{tabular}
  \caption{Algebraic structure of the spin states for $M=3$. 
  \label{tab:spinwaves-alg}}
  \setlength{\tabcolsep}{6pt}
  \renewcommand{\arraystretch}{1.3}
  \begin{tabular}{|c|c|c|c|c|c|} \hline
  state & $\Okin$ & $O_z$ & $O_1$ & $H_\Heis$ & $I_S^z$\\ \hline
    $\ket{Z^\ka}$ & 0 & $+\frac{3}{4}$ & $+\frac{3}{4}$ & 0 & 0 \\ \hline 
    $\ket{S^\ka_0}$ & 1 & $-\frac{1}{4}$ & $+\frac{3}{4}$ & 0 & 0 \\ \hline 
    $\ket{S^\ka_\pm}$ & $-\frac{1}{2}$ & $-\frac{1}{4}$ & $-\frac{3}{4}$ & $- 6J^2/U$ & $\pm \sqrt{3}\, {\ka}$ \\  \hline 
  \end{tabular}
  \caption{Kinetic parameters of the spin states for $M=3$, {$\ka = \pm$}. {(The splitting $O_1 = \Okin + O_z$ is introduced in \eqref{ES+OS} in App.~\ref{app:Spin-waves}.)}
  \label{tab:spinwaves-kin}}
\end{table}

\subsection{Magnons} \label{sec:Trimer-Magnons}

{
For $U \gg J$, there will be also $8$ energetically low lying eigenstates $\ket{\psi_i}$ of the full Hubbard system, close (of the order of $J/U$) to the eigenstates of corresponding Heisenberg Hamiltonian,
which} can be grouped as follows:

\tbul {four} non--dynamic fully spin polarized states $\ket{\psi^\ka_{Z}} = \ket{Z^\ka}$ 
{and $\ket{\psi^\ka_0} = \ket{S^\ka_0}$}
for $\ka = \pm$ with maximal spin $S=\frac{3}{2}$ {and ferromagnetic order} 
which are identical {with the Heisenberg eigenstates discussed above and have 
total energy $\ev{H_S} = H_0$,}

\tbul two pairs of states $\ket{\psi^\ka_\pm} \approx \ket{S^\ka_\pm}$ 
with wavenumbers $k= \pm  k_0$ and $S=\frac{1}{2}$ being close to the spin waves with small mixtures of doublon--holon pairs of the form $\ket{DH1_\pm}$ (which means double occupation (D), zero occupation (H) and single occupation (1) distributed over different sites) such that, symbolically, $\ket{\psi^\ka_\pm} = \al \ket{S^\ka_\pm} + \be \ket{DH1^\ka_\pm}$ with the amplitudes $\al \approx 1$ and $\be \approx J/U$  (the details are given in App. \ref{app:DH1}).
Their energy is, cf. \eqref{E-magnons},
\begin{equation}
  E_{\pm} = \ev{H_S}_{\psi^\ka_\pm} = -\frac{3}{2}U - \frac{6 J^2}{U}
  + \Ocal{\frac{J^3}{U^2}}
\end{equation}
and lies below $E_{0} = \ev{H_S}_{\psi^\ka_0}$ as the kinetic energy contributes negatively.
Their spin current expectation values are, cf.~\eqref{IS-magnons},
\begin{equation}
  \ev{I_S^z}_{\psi^\ka_\pm}
  \approx \pm \sqrt{3} \left(1-\frac{3J^2}{U^2}\right)^2
\end{equation}
and the spin order is, cf. \eqref{O1-magnons},
\begin{equation}
  \ev{O_1}_{\psi^\ka_\pm}
  \approx - \frac{3}{4} \left(1-\frac{3J^2}{U^2}\right)^2.
\end{equation}
These values of energy and order can be seen in Fig. \ref{fig:E+O-LateRelax-Non}.

An important observation is that {since} the wavefunctions and the energies of the Heisenberg spin waves {$\ket{S^\ka_s}$} and the corresponding Hubbard magnons {$\ket{\psi^\ka_s}$} are close to each other (up to a power of $J/U$),
the action of the Lindblad operators which is based on the (local or global) energy eigenstates remains almost unchanged under the UDL--transformation and we phenomenologically expect the same behavior:
global Lindblads tend to minimize the global energy, local Lindblads the local on--site energy (cf. Sec. \ref{sec:Late-Relax}).


{In the sections below, we choose particular initial states and conditions 
and discuss the relaxation of different observables by solving numerically 
\eqref{Lind-A} with different Lindblad jump operators.
}

\subsection{Persistent spin current at $T=0$}

In order to demonstrate the relaxation to non--trivial magnonic final states we first look closer at the initial states which already carry a non--vanishing spin current $I_S$.
To observe the stationarity of the spin current flow we choose an initial state consisting purely of the spin wave modes, which are supposed to be close to the exact magnonic eigenstates, as discussed above. 
The picture becomes more dynamical when two waves with different energies are superimposed, e.g. $\ket{S^\ka_0}$ and $\ket{S^\ka_+}$ (we fix $\ka=+$ or $-$ and skip the index because both behave identically)
into 
$\ket{\psi} = (\ket{S_+} - \ket{S_0})/\sqrt{2}$
and $\rho(0) = \ketbra{\psi}{\psi}$,
because the relative oscillations between them lead to an interference pattern propagating across the lattice, cf. Fig. \ref{fig:spincur_density}.

\begin{figure}[t]
  \includegraphics[width=0.9\linewidth]{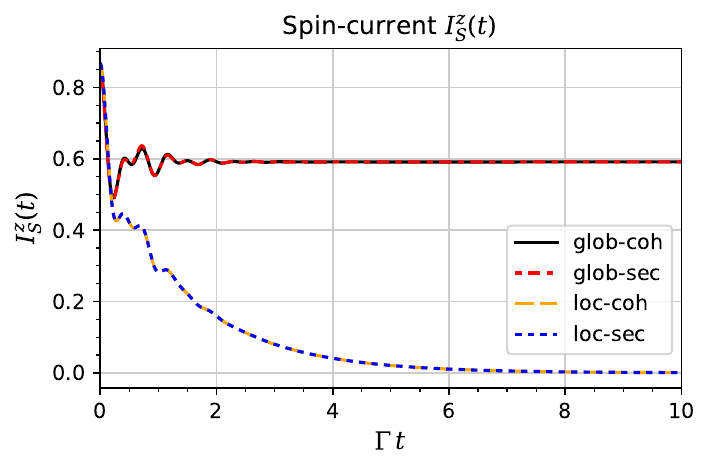}
  \includegraphics[width=\linewidth]{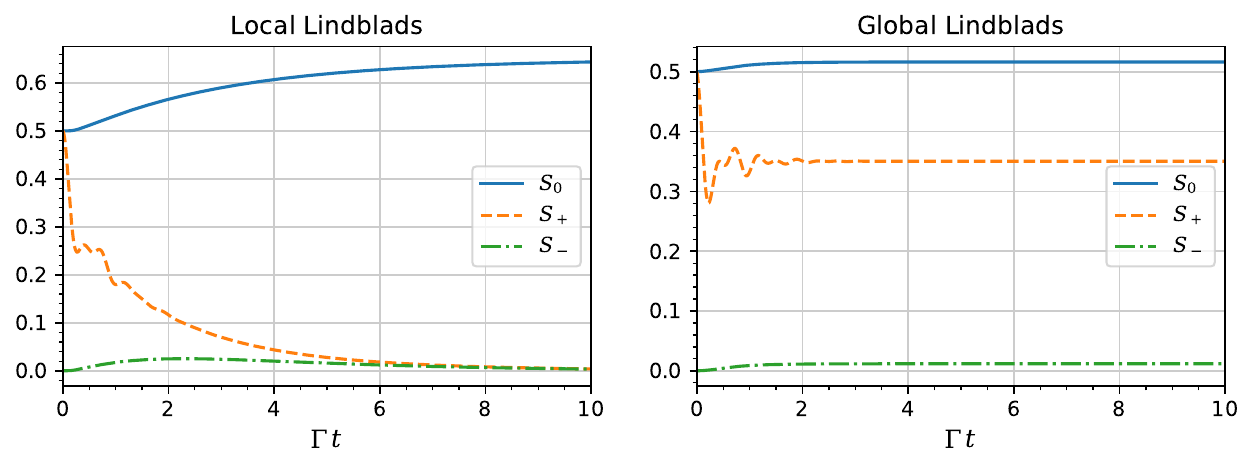}
  \caption{Spin current $I_S^z(t)$ (top) and projections onto the spin waves $\ket{S_s}$ with $s\in\{0,\pm 1\}$ (bottom)
  for the initial state 
  $\ket{\psi} = (\ket{S_+} - \ket{S_0})/\sqrt{2}$. 
  The spin current persists for global but decays for local Lindblad operators (secular and coherent almost identical).
  \label{fig:pers-cur_spin-cur}
  With global Lindblads all components relax only slightly and quickly settle to final values whereas with local Lindblads only the $\ket{S_0}$ component remains non--zero while $\ket{S_\pm}$ carrying spin current decay to zero (secular and coherent almost identical).
  The single spin excitation states $\ket{S_s}$ have a large overlap with the corresponding magnons $\ket{\psi_s}$ and therefore can be used as an approximate projection basis.
  (Other parameters as in Fig. \ref{fig:N}.)
  \label{fig:pers-cur_proj-spinwaves}}
\end{figure}

As we see in Fig. \ref{fig:pers-cur_spin-cur}, the spin current ${I^z_S(t)}$ is present from the beginning. 
Initially, it falls because of the decay of the spin wave states (which are not exact eigenstates) but then stabilizes at around $2/3$ of its initial value when we use the global Lindblad operators (the choice of coherent or secular leads here to the same result). 
It is an interesting observation that the spin current does not decay further despite coupling to the local baths. The reason is that at $T=0$ the further decay of energetically low lying states with the baths is blocked. 
On the other hand, local Lindblad operators do not respect the global energy scale and enable local  
exchange of electrons between the sites and their respective baths as long as the local occupation differs from one electron per site (even at $T=0$). 
Since the magnonic eigenstates $\ket{\psi_\pm}$ are close to the spin waves $\ket{S_\pm}$ but have a small ($\sim J/U$) mixture of doublons and holons they slowly decay via the system--bath channels and the system is driven towards the spin wave with vanishing wavenumber $\ket{\psi_0} = \ket{S_0}$ which, in contrary, does not contain any doublons nor holons and thus does not decay any further. 

Since we claim that for $\Ga \ll J$ the global Lindblad operators provide a more accurate description of the dynamics than the local ones we expect the persistent spin current to be a generic final state for the system in this regime. 
If more than one magnonic states with different energies are involved, we also expect periodic oscillations of the local spins%
\footnote{For any two decoherence--free states $\ket{\phi}, \ket{\chi}$ which are annihilated by all Lindblad operators,   
$L_\al \ket{\phi} = L_\al \ket{\chi} = 0$ for all $\al$, 
their coherence is also decoherence--free, $\L \ketbra{\phi}{\chi} = 0$.
This happens for all of the eight states listed in Tab. \ref{tab:spinwaves-alg} and \ref{tab:spinwaves-kin}.
All pairs with different energies lead, via their coherences, 
to oscillations in some observables, e.g. the local spin--$z$.}.
This is only observed in relaxation processes within the global coherent approximation since in the global secular approach the coherences between states with different energies do not get excited at all%
\footnote{For $M=2$  the oscillation of $\D S^z = S_1^z - S_2^z$ because of the persistent coherences between triplet and singlet states $\ketbra{T_i}{S}$) have been observed in \cite{B07-Dimer}.
}. 

\subsection{Relaxation to persistent spin current at $T=0$}

\begin{figure}[t]
  \includegraphics[width=0.9\linewidth]{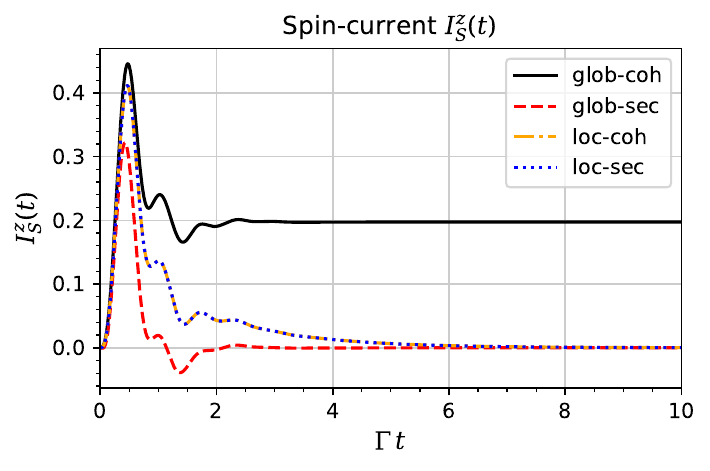}
  \includegraphics[width=0.9\linewidth]{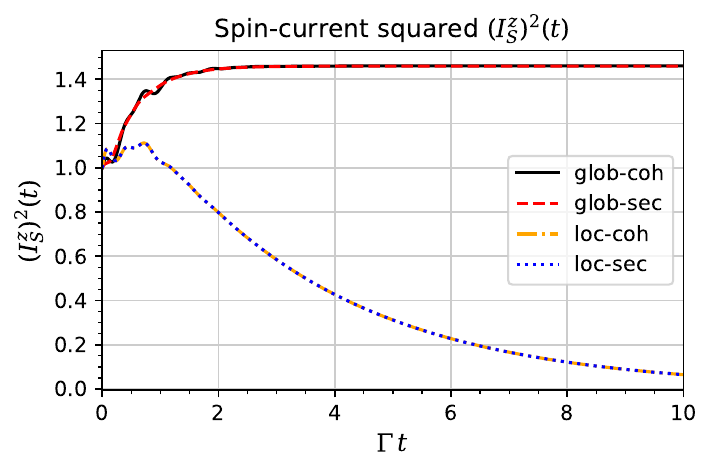}
  \caption{Relaxation to the persistent spin current from the initial state $\ket{\psi} = \ket{\su,0,\sd}$. Top: Spin current $I_S^z(t)$ decays for both local coherent and secular Lindblads and for the global secular Lindblads while it persists only for the global coherent Lindblads. 
  Bottom: The spin current squared $(I_S^z)^2(t)$ reveals that the spin current is non--zero in both global cases, coherent and secular. but in the secular case it is left--right symmetric with  both magnons $\ket{\psi_\pm}$ identically excited such that their contributions to the net spin current $I_S^z(t)$ cancel out. 
  (Other parameters as in Fig. \ref{fig:N}.)
  \label{fig:relax-to-spin-cur}
  }
\end{figure}

\begin{figure*}[t]
  \includegraphics[width=1.0\linewidth]{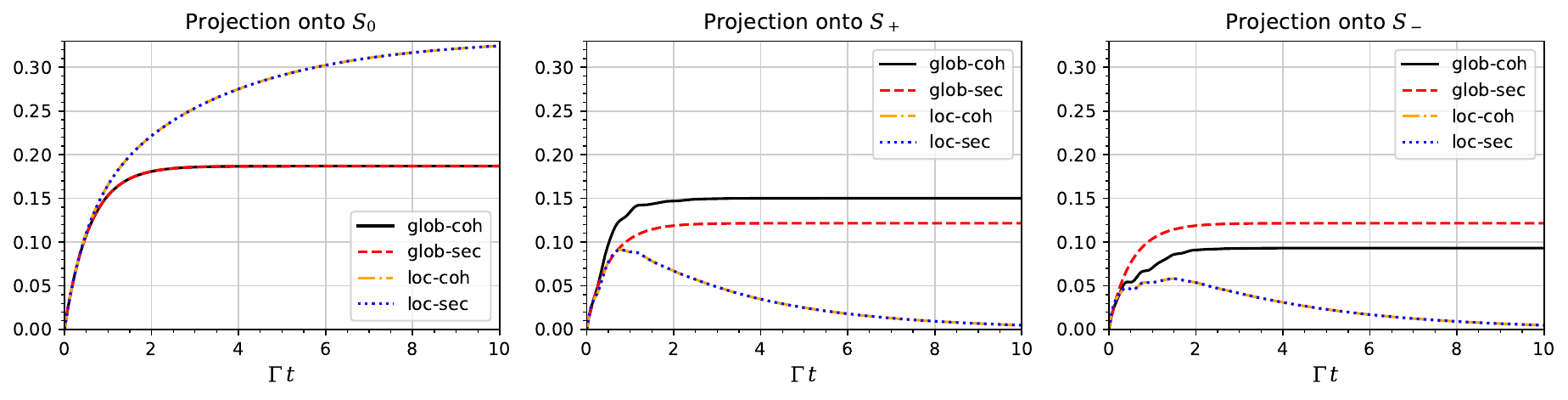}
  \caption{Projections onto the spin waves $\ket{S_s}$ with $s\in\{0,\pm \}$ for the initial state 
  $\ket{\psi} = \ket{\su,0,\sd}$. 
  With both local Lindblads only the $\ket{S_0}$ component remains non--zero while $\ket{S_\pm}$ carrying spin current decay to zero (secular and coherent almost identical).
  With the global secular Lindblads both ground state magnons $\ket{S_\pm}$ must get equally excited thus leading to left--right symmety and net zero spin current $I_S^z(t)$. 
  Global coherent Lindblads do not enforce the left--right symmetry in the excitation of $\ket{S_\pm}$ and hence a non--vanishing directed spin current $I_S^z(t)$ may persist in the system. 
  (Other parameters as in Fig. \ref{fig:N}.)
  \label{fig:relax-to-spin-cur_proj}
  }
\end{figure*}

In order to show that the relaxation to the persistent spin current is a generic process for this system we begin with an excited state with no overlap with any spin wave or magnon states, $\rho(0) = \ketbra{\psi}{\psi}$ with $\ket{\psi} = \ket{\su,0,\sd}$.
For $J \gg \Gamma$, first the spin polarized current $J^z_S$ starts to flow because of the tunneling of both electrons with different spins from the opposite directions into the empty middle site (with the Coulomb blockade acting in the other direction over the periodic bourndary). 

Even for the isolated system ($\Ga = 0$), the spin polarized currents ${\v J_S(t)}$ are not conserved when $U \neq 0$.
For the initial state $\ket{\su,0,\sd}$, which breaks the (cyclic) left-right lattice symmetry, the current ${\v J_S(t)}$ oscillates ---
it changes direction whenever the two opposite spins ``collide'' (owing to the strong Coulomb interaction) and change direction.
When $\Ga > 0$,
the third electron tunnels into the system 
and 
the system state decays towards the Heisenberg sector with local half filling. 
The spin polarized current ${\v J_S(t)}$ turns into the pure spin current ${\v I_S(t)}$ which is conserved. 
In consequence, we observe the decay of the oscillating current ${J_S^z(t)}$ into the conserved current ${I_S^z(t)}$ whose final value will strongly depend on the details of the dynamics, in particular on the relation between $\Ga$ and $J$ which determine the decay rate and the oscillation frequency, respectively.

Owing to the interaction with the baths, the initially pure state decays,
by the injection of the third electron with spin $\su$ or $\sd$, to an incoherent sum of two sectors with different total spin--$z$, $S^z$ (incoherent because electrons with different spins tunnel into the system via different channels described by separate Lindblads, cf. \eqref{LL-sec-Lind}).
Each of the two sectors is spanned by the magnons
which lie deep in the energy spectrum and at $T=0$ do not couple to the baths and stay stationary. 
Here, we observe a strong discrepancy between the global coherent and secular approximations. 
The reason is that global Lindblad operators are defined w.r.t. the eigenstates of the system Hamiltonian $H_S$. 
In the secular approximation, the degenerate pair of left and right moving magnons $\ket{\psi_\pm}$ with wavenumbers $k_\pm = \pm k_0$
combine incoherently with the non--moving magnon with the wavenumber $k_0 = 0$
and become immediately static while in the coherent approximation their sum is always coherent. Since the energies satisfy $E_0 \neq E_+ = E_-$ the states, in the latter case, begin to oscillate thus leading to a periodic dynamics of the local spin on the lattice (cf. Fig \ref{fig:spincur_density}).
In contrast, in the secular case, the spin after relaxation is always fully delocalized.
This exposes a serious limitation of the secular approximation: although the third electron tunnels into the system via a single channel from a single bath (the other baths could even be non--existent at that moment) it is immediately delocalized and present in all sites with the same probability.
This contradicts the finite propagation speed in such systems.
It is consistent with the previous observations of the discrepancy between the global coherent and secular approaches in the dimer \cite{B07-Dimer} in which similar oscillations between the singlet and triplet states playing there the analogue role to the spin waves have been seen in the coherent approximation.

As we can see in Fig. \ref{fig:relax-to-spin-cur}, the spin current always decays to zero for the local Lindblads which maximize the spin and spin order in the system and thus prefer ${\ket{\psi_0}}$ ($S=\frac{3}{2}$) over $\ket{\psi_\pm}$ ($S=\frac{1}{2}$).
Both global Lindblads lead to non--vanishing spin currents (confirmed by ${(I_S^z)^2(t)} > 0$) but in the secular case the net current ${\v I_S(t)}$ always vanishes. 
The reason is not in the different treatment of the two magnonic ground states $\ket{\psi_\pm}$ because, as they have equal energies, both types of global Lindblads treat them coherently. 
The difference is in the treatment of states $\ket{\psi_i}$ (in this example, occupied by two electrons) from which the system can decay into $\ket{\psi_\pm}$. 
Since they consist of eigenmodes with different energies each separate transition is incoherent in the secular scheme, resulting always in identical amplitudes of $\ket{\psi_\pm}$. In contrast, in the coherent scheme, a coherent transition from $\ket{\psi_i}$ may lead to different final amplitudes of $\ket{\psi_\pm}$, cf. Fig. \ref{fig:relax-to-spin-cur_proj}.

This reveals the second strong discrepancy between the coherent and the secular approximation. 
While within the coherent approach, the modes $\ket{\psi_\pm}$ carrying the right and left propagating spin currents can be excited independently, in the secular approach their amplitudes are always equal. 
In consequence, the coherent approach supports non--zero directed persistent spin currents while the secular {does} not{, in full agreement with the statements of \cite{Breuer-CohLind}}. 
The net spin current is potentially a measurable observable which can give the answer which of the Lindblad approaches is the closest one to reality. 

\subsection{Relaxation to spin current for $T \gtrsim 0$}

\begin{figure}[t]
  \includegraphics[height=5.5cm]{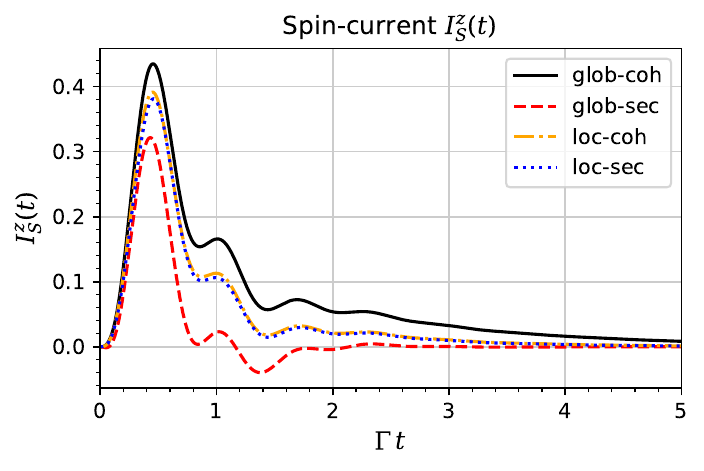} 
  \includegraphics[height=5.5cm]{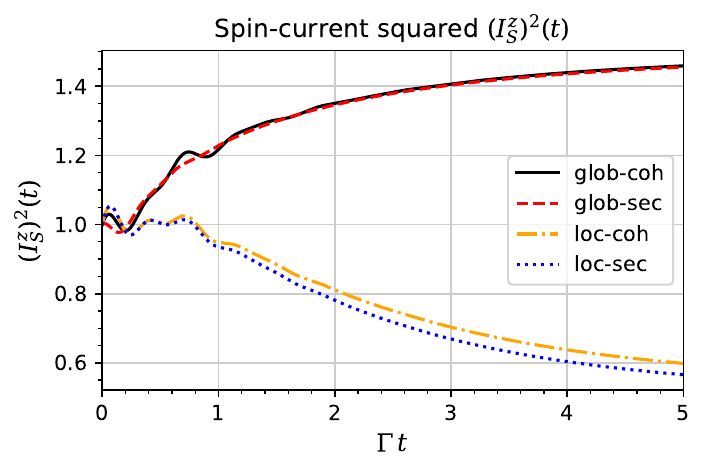} 
  \caption{Relaxation to persistent spin current at small temperature $T=2\,\Ga$ from the initial state $\ket{\psi} = \ket{\su,0,\sd}$. 
  Top: Spin current $I_S^z(t)$ decays for all Lindblads. 
  Bottom: The spin current squared $(I_S^z)^2(t)$ reveals that the spin current is non--zero in both global cases, coherent and secular, however it is left--right symmetric with both magnons $\ket{\psi_\pm}$ identically excited such that their contributions to the net spin current $I_S^z(t)$ cancel out. 
  (Other parameters as in Fig.~\ref{fig:N}.)
  \label{fig:relax-to-spin-cur+T}
  }
\end{figure}

Finally, we want to understand how universal the relaxation to the magnonic ground states $\ket{\psi_\pm}$ carrying spin currents is. 
We expect that{, with global Lindblads} at small but positive temperature $T \gtrsim 0$, the system will globally relax to the Gibbs state, being close to the above mentioned ground states {$\ket{\psi_\pm}$}. 
However, it turns out that at $T>0$, the occupations become left--right symmetric (because of same energy), with both magnons $\ket{\psi_\pm}$ equally excited, and incoherent (because of the mixed Gibbs state) what leads to a vanishing net current, ${I_S^z(t) \ra 0}$, as both spin current contributions, left and right, cancel each other out, cf. Fig. \ref{fig:relax-to-spin-cur+T} (top). 
The spin current operator squared, ${(I_S^z)^2(t) \equiv \ev{(I_S^z)^2}_t}$, related also to the fluctuations, confirms that indeed spin currents are present in the system, ${(I_S^z)^2(t)} > 0$, cf. Fig. \ref{fig:relax-to-spin-cur+T} (bottom). 
We observe no oscillations between states at different energies $E_0 \neq E_\pm$ since at $T>0$ the system tends to a mixed state with no coherences.

\subsection{Relaxation to spin current with magnetic flux and $T \gtrsim 0$}

\begin{figure}[t]
  \includegraphics[height=5.5cm]{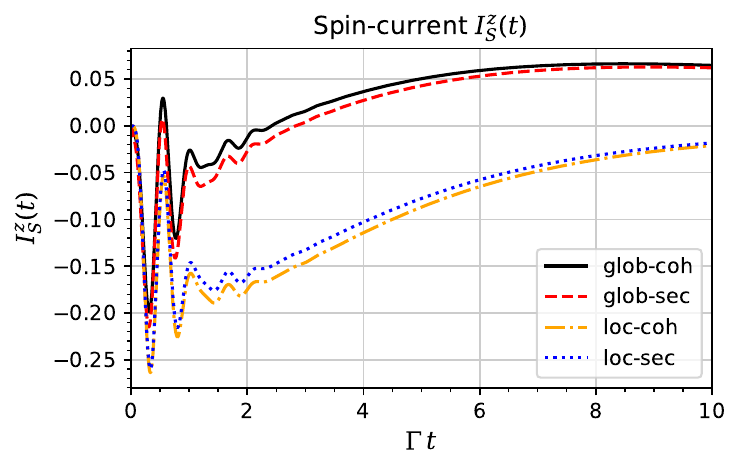} 
  \includegraphics[height=5.5cm]{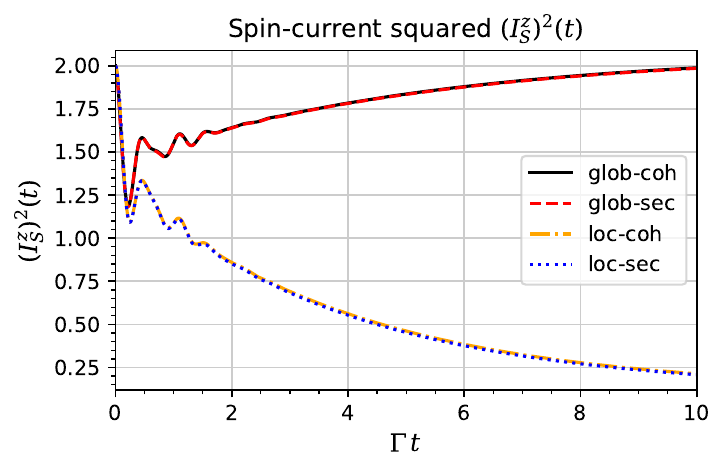} 
  \caption{Relaxation to persistent spin current at small temperature $T=2\Ga$ with magnetic flux $B=0.5\,\Ga$ from the initial state $\ket{\psi} = \ket{\su,\su,\sd}$. 
  Spin current $I_S(t)$ decays to zero for both local Lindblads but stabilizes at a persistent value for both global Lindblads. 
  Because of the left--right (chiral) symmetry breaking by the magnetic flux $B$ the degeneracy between the two magnons $\ket{\psi_\pm}$ is lifted which facilitates a directed persistent spin current $I_S(t)$.  
  (Other parameters as in Fig.~\ref{fig:N}.)
  \label{fig:relax-to-spin-cur+T+B}
  }
\end{figure}

In order to break the left--right symmetry, we add to the system Hamiltonian $H_S$ a small non--zero magnetic flux {$F_B$}, corresponding formally to the field strength $B$, perpendicular to the trimer plane. 
{To achieve this,} we multiply the hopping terms $J$ {in \eqref{Hubbard}} with the Peierls phases \cite{Peierls-Phase, QuantumRingsForBeginners} $\exp(i \v A \v d)$%
\footnote{The magnetic flux $F_B = B A_\D$ flows through the area $A_\D = \sqrt{3} d^2/4$ of the equilateral triangle with sides $d$ build by the trimer. From the Stokes theorem, $3 |\v A| d = B A_\D$, and hence the vector potential $|\v A| = \sqrt{3} B d/12 $. Since the units are arbitrary in our example we choose $d = 4 \sqrt{3}$ so that $B = |\v A|$.
}.

In Fig.~\ref{fig:relax-to-spin-cur+T+B} we can see that local Lindlads again destroy the spin current ${I_S^z(t)}$ (also its square vanishes). 
However, the magnetic field $B$ breaks the left--right (chiral) symmetry and thus lifts the degeneracy between the two magnons $\ket{\psi_\pm}$ such that only one becomes a unique ground state and acts as a global attractor of the dynamics. 
Thus, we observe relaxation to a directed persistent spin current ${I_S^z(t)}>0$ for both global Lindblad types.
 
\section{Discussion and outlook}

In summary, we have studied different stages of relaxation in the strongly interacting Fermi--Hubbard system (quantum sites) coupled to cold fermionic baths (environment). 
The early fast relaxation of the global energy and global particle number proceeds universally with the rate $2\, \Ga$
and brings the system to the low--energy sector, well described by the Heisenberg model. 

Depending on the coupling parameter regime, $\Ga$ vs $J$, and on the corresponding approximation scheme, only the late relaxation proceeds qualitatively differently, leading the system to the lowest global energy with maximal antiferromagnetic order and minimal spin for $\Ga \ll J$ or to the lowest local energy with maximal ferromagenetic order and maximal spin for $J \ll \Ga$.
In the $J \ll \Ga$ regime, approximated by local Lindblad operators, the late relaxation of the system energy and spin order proceeds slowly, at the rate $\sim \Ga J^2 / U^2$, even at zero temperature, $T=0$.
In the $\Ga \ll J$ regime, approximated by global Lindblad operators, the late relaxation depends on the temperature of the environment and at low temperatures, $T\approx 0$, proceeds ultra slowly at the rate  $\sim \Ga \exp(- U/(2T))$.
The fact that different approximations may lead to different decay mechanisms and final states of different types
should be a warning for a too easy justification of the chosen approximation scheme.

In the special case of a trimer (ring--shaped three sites, $M=3$) the spin frustration prohibits the perfect antiferromagnetic order and thus leads to low energetic spin waves (magnons) which play the role of the (degenerate) ground states and final states of evolution (global attractors).
Consequently, in the regime with $\Ga \ll J$, the system universally settles down to the magnonic final states which carry a persistent spin current.
This current remains stable and decoherence free despite the  interaction with the environment.
We expect similar behavior in larger systems with odd numbers of sites $M$, however, 
this question requires further investigation.

This observation generalizes from the singular case of zero temperature $T=0$ (in which some additional trapping in other low--energy states also takes place) to small positive $T>0$.
Since the thermal loss of coherence slowly leads the system to a fully mixed state of spin current modes propagating in the opposite directions, and hence to a zero net spin current $I_S^z$, the addition of a small magnetic flux lifts that degeneracy and prefers one specific persistent current flow direction.

It remains an open question, how the interplay between the coupling strengths $\Ga$ (system--bath) and $J$ (internal system) determines the proper choice of the approximate description based on Lindblad operators. Here, we studied the two extreme regimes, leading to global or local Lindblads, 
but it is an important further research direction to better understand the regime between the two extremes.

The statements about the persistent current in the final states should also hold true for the Redfield equation%
\footnote{Without the imaginary terms in \eqref{F+iB}, 
whenever it yields mathematically acceptable results and is compatible with the thermalization. 
In that case, all factors in \eqref{Redfield-matrix} for processes leading out of the ground states vanish, too, due to $f_+(\D E)=0$ for $\D E>0$.}. 
However, the relaxation towards them may proceed differently. 

Including higher orders in $\Ga$ would necessarily lead to the on--site spin--flip processes and, in consequence, to a contribution with a symmetric left--right distribution of the spin currents, 
similar to the situation with $T>0$.

The system can be realized by tunnel--coupled quantum dots \cite{TrimerQDotsExp, ALudwig-QDot-Array, Ensslin+Kurzmann-QDots+BilGraphene}. 
{Recent progress in gate defined quantum dots regarding state preparation, readout and system parameter manipulation \cite{Haug-QD-ChargeMess-3-4, QD-Plaquette-Nagaoka, QD-Plaquette-QSim} enables at least indirect detection of relaxation to spin waves by fast decoupling of the dots ($\Ga, J \ra 0$) and measurement of the double and null occupations ($\NDH$) in each dot. Their equal presence in all dots would suggest propagating (ground state) magnons carrying spin current while their absence would indicate non--propagating state with ferromagnetic spin order. 
Furthermore, the dependence of $\NDH$ on the magnetic flux through the trimer can help to determine the spin current direction.}

An alternative option could be a realization of the trimer in optical lattice systems which  offer the advantage of highly tunable parameters and precise measurements \cite{Bloch-SpinCurInOL, Amico+Osterloh-Ring-OL, Amico-PersistenCurrents-OL, PersCur}. 
{Setting the effective on--site interaction $U \ra 0$ after the relaxation phase induces doublon--holon oscillations in all sites only if the system was in a magnonic ground state carrying non--zero spin current.}

It will be valuable to study larger systems and answer the question whether the spin current modes are generally important final states in relaxation processes driven by the environment \cite{Chernyshev2009SpinWavesTriLat}.
Also the preference of antiferromagnetic order and its relation to excitation of spin wave modes (magnons) in systems with odd and even numbers of sites, supporting the spin frustration or not, is an interesting open question \cite{Wietek2021CompetingOrdersTrigLat}.

{
Finally, bosonic instead of fermionic baths can be considered \cite{queisser2019a, QDots+reservoir} which do not exchange particles with the system but only energy or spin. In \cite{queisser2019a} also two relaxation stages, an early local and a late global one, have been observed, however, based on different physical signatures.
}
 
\acknowledgements

The authors thank Lukas Litzba for fruitful discussions and valuable feedback on the manuscript.
We gratefully acknowledge funding by the Deutsche Forschungs\--gemeinschaft (DFG, German Research Foundation) --- Project 278162697 --- SFB 1242.
We thank the developers of QuTiP \cite{QuTiP} which was used in the numerical calculations.

\appendix

\section{Redfield and Lindblad equations} \label{app:Lindblad}

\subsection{From microscopic to Redfield description} \label{app:Lindblad-Redfield}

By applying the Born and the first Markov approximations, 
but before the second Markov approximation ({equivalent to taking the limit} $t_0\ra -\infty$), 
and by tracing out the bath degrees of freedom in the von Neumann equation, 
$ \p_t\, \rho(t) = -i\, [H, \rho(t)] $,
expanded to the lowest non--vanishing order in the system--bath couplings, $\Ocal{|\ga_{m,s,k}|^2}$,
we obtain the (first) Redfield equation \cite{Redfield, redfield1965a}
\begin{equation} \label{Redfield-app}
   \p_t\, \rho(t) = -i\, [H_S, \rho(t)] + \L_t\, \rho(t).
\end{equation}
The superoperator%
\footnote{We refer to the operators ``superoperators'' when they directly act on the full Liouville space.} 
$\L_t$ can, in general, be not posi\-tivity preserving (short \textit{not positive})%
\footnote{The superoperator $\L$ is positivity preserving 
when the map $\mathcal{U}_t \equiv e^{\L t}$ preserves the positivity of the density matrix $\rho(t) = \mathcal{U}_t \rho(0)$ for all $t \geq t_0$. It is also required that $\L$ preserves the trace of $\rho$.}, 
which is related to the derivation of the Redfield equation at the lowest order in the system--environment coupling strength, cf. \cite{StayingPositive, Spohn-PositivityPreserving, Breuer, GSchaller-Review-NQSys}.
The superoperator $\L_t$
can be split into two parts, $\L_t = -i [\d H_S(t), \cdot] + \t \L_t$, of which the first can be included in the ``renormalized'' (or ``Lamb--shifted'') Hermitian, possibly time--dependent, Hamiltonian 
\begin{equation} \label{H-ren}
\begin{split}
	&\t H_S(t) = H_S + \d H_S(t) = H_S \\
	&- i\!\!\!\! \sum_{\substack{m,s,\al \\ \D E, \D E'}} \frac{F_\al(t,\D E) - \overline{F_\al(t,\D E')}}{4} 
	K^\al_{m,s}(\D E')^\dagger K^\al_{m,s}(\D E)
\end{split}
\end{equation}
with 
\begin{equation}
   K^\al_{m,s}(\D E) = \sum_{k,l} \d_{E_k-E_l,\al \D E} \ketbra{E_k}{E_k} c^\al_{m,s} \ketbra{E_l}{E_l}
\end{equation}
where $c_{m,s}^+ = c\s_{m,s}, c_{m,s}^- = c_{m,s}$
and
\begin{multline} \label{F(t)}
  F^\al_{m,s}(t,\D E) =
  2 \int_0^{t-t_0}\text{d}\tau \sumint_k  
  |\ga_{m,s,k}|^2 f_\al(\om)\, e^{i \al (\eps_k - \D E) \tau} \\
  = \int_0^{t-t_0}{d}\tau\int_{-\infty}^\infty\frac{{d}\om}{\pi} \Ga_{m,s}(\om) f_\al(\om)\, e^{i \al (\om - \D E) \tau}
\end{multline}
where in the last step the distribution $\Ga_{m,s}(\omega) = 2 \pi {\int \hspace{-8.5pt} \Sigma}_k |\ga_{m,s,k}|^2 \d(\om-\eps_k)$ has been introduced.
In the following, we assume that the bath's spectrum is dense, replace the sum with an integral and assume its density is uniform, i.e. the \textit{wide-band limit}, 
$\Ga_{m,s}(\omega) \ra \Ga_{m,s}$.
{Above,} $\D E$ 
refer to all possible differences of eigenenergies of the system Hamiltonian $H_S$ while 
$\al$ refers to creation ($+$) and annihilation ($-$) processes. 

That splitting brings the Redfield equation \eqref{Redfield-app} to the form
\begin{equation} \label{Redfield-mod}
	\p_t \rho = -i [\t H_S, \rho] + \t \L_t \, \rho
\end{equation} 
with the new Liouville superoperator
\begin{equation} \label{LL}
	\t \L_t \rho = 
	\sum_{\substack{m,s,\al \\ \D E, \D E'}} M^\al_{m,s}(t, \D E, \D E')\, \L_{m,s}^\al(\D E, \D E')\, \rho,
\end{equation}
including the coefficients
\begin{equation} \label{M=F+F}
	M^\al_{m,s}(t, \D E, \D E') = \frac{F^\al_{m,s}(t,\D E) + \overline{F^\al_{m,s}(t,\D E')}}{2}
\end{equation}
and the superoperators
\begin{multline} 
	\L_{m,s}^\al(\D E, \D E') = 
	K^\al_{m,s}(\D E)\, \rho\, K^\al_{m,s}(\D E')^\dagger \\
	 - \frac{1}{2} \left\{ K^\al_{m,s}(\D E')^\dagger K^\al_{m,s}(\D E), \rho \right\}.
\end{multline}
The objects $K^\al_{m,s}(\D E)$ and $\L_{m,s}^\al(\D E, \D E')$ describe {elementary} processes occurring in the system while the objects $F^\al_{m,s}(t,\D E)$ and $M^\al_{m,s}(t, \D E, \D E')$ their intensities and all carry the residual information about the properties of the baths and their couplings to the system.

The total Liouville superoperator $\t \L_t$ is still not positivity preserving, however the {individual} superoperators $\L_{m,s}^\al(\D E, \D E')$ in the sum \eqref{LL} can be diagonalized in the energy basis to become positive Lindblad superoperators.
This suggests that the origin of non--positivity is solely related to the structure of the coefficients $M^\al_{m,s}(t, \D E, \D E')$.
Below, we apply further steps in order to obtain modified coefficients which restore the overall positivity. 

By applying the second Markov assumption and integrating over an infinite history, which means taking the integration limit $t_0 \ra -\infty$ in \eqref{F(t)}, we obtain static coefficients
\begin{equation} \label{F+iB}
  F^\al_{m,s}(t,\D E) = \Ga_{m,s}[ f_\al(\D E) + i B_\al(\D E) ]
\end{equation}
where the real part $f_\al(\D E)$,
\begin{multline}
  \Re [ F^\pm_{m,s}(t,\D E) / \Ga_{m,s} ] = \frac{1}{2} \mp \int_0^{\infty} T \frac{\sin(\D E \tau)}{\sinh(\pi T \tau)} d\tau \\
  = \frac{1}{2} \mp \frac{1}{2}\tanh\left( \frac{\D E}{2 T} \right) 
  = f(\pm \D E) \equiv f_\pm(\D E),
\end{multline}
is obtained directly from the Fermi--Dirac distribution of  the baths
and the the imaginary part, $B_\al(\D E)$, was defined in \cite{B07-Dimer}. 
Its properties were discussed in more detail in \cite{EK-PhD} but it will play no role in the present context%
\footnote{The differences, entering  \eqref{M=F+F}, can be estimated by $B(\D E)-B(\D E') = \Ocal{{\Ga J}/{U}}$ and are small in our case.}. 

The Redfield equation is generally valid in the regime when the system--bath couplings are small relative to the energy differences which occur during the tunneling, i.e. $\Ga_{m,s} \ll |\D E| \approx U$.
In the following, we will neglect the renormalization effects in the Hamiltonian \eqref{H-ren} as they usually do not influence the dynamics significantly. 
We will also, for better readability, assume spin and site--independent tunneling rates, 
i.e. $\Ga_{m,s} \ra \Ga$.
The generalization to site dependent couplings $\Ga_{m,s}$ is straightforward.

\subsection{From Redfield to Lindblad equation} \label{app:Lindblad-Lindblad}

In the derivation of the above Redfield equation, at the lowest order in the system--environment coupling strength, 
the positivity of the evolution generator gets lost. 
This has the consequence that probabilities may become negative or grow unboundedly, cf. \cite{B07-Dimer, EK-PhD, LL-MTh}.
Here, we want to correct it in order to restore the positivity and thus obtain a Lindblad master equation.
Even if it is not completely clear whether the Lindblad equation generally gives results closer to the exact solution than Redfield, and there exists valid criticism \cite{Hatrmann+Strunz-RedfieldNonpositivity,  Nathan-CohLindDerivation, Purkayastha-LindbladCritics}, it provides acceptable {long--term} evolution and final states where Redfield usually fails \cite{EK-PhD}.

\subsubsection{Global secular approximation}

For pairs of energy differences in the system, $\D E$ and $\D E'$, let us consider the blocks
\begin{widetext}
\begin{equation} \label{Redfield-matrix}
\begin{split}
  \M^\pm_{\D E, \D E'}  &=
  \begin{pmatrix}
    M^\pm(\D E, \D E) & M^\pm(\D E, \D E') \\
    M^\pm(\D E', \D E) & M^\pm(\D E', \D E')
  \end{pmatrix} \\
  &= \Ga
  \begin{pmatrix}
    f_\pm(\D E) & 
    \frac{f_\pm(\D E) + f_\pm(\D E')}{2} + i \frac{B_\pm(\D E) - B_\pm(\D E')}{2} \\
    \frac{f_\pm(\D E') + f_\pm(\D E)}{2} - i \frac{B_\pm(\D E) - B_\pm(\D E')}{2} & 
    f_\pm(\D E')
  \end{pmatrix}
\end{split}
\end{equation}
\end{widetext}
which are not positive definite for $\D E \neq \D E'$, having one positive and one negative eigenvalue.
This non--positivity is responsible for the non--positivity of the total Liouville operator in \eqref{LL}.
The off--diagonal elements 
{are responsible for the excitation of}
coherences between states with different energies (in the Liouville space), which oscillate in time. 
The secular approximation averages out the oscillations and effectively removes the off--diagonal terms and brings the above blocks to the diagonal form
\begin{align}
  \M^{\pm,\text{sec}}_{\D E, \D E'}  = \Ga
  \begin{pmatrix}
    f_\pm(\D E) & 0 \\
    0 & f_\pm(\D E')
  \end{pmatrix}
\end{align}
which are obviously positive semidefinite
{(in the degenerate case, $\D E = \D E'$, the matrix $\M^{\pm,\text{sec}}_{\D E, \D E}$ is not diagonal but contains four identical entries, $f_\pm(\D E)$, and hence is also positive semidefinite)}. 
It is equivalent to the replacement of the coefficients $M^\pm$ with
\begin{align}
    M^{\pm,\text{sec}}(\D E, \D E') = \delta_{\D E, \D E'} M^\pm(\D E, \D E').
\end{align}
In consequence, the Liouville superoperator \eqref{LL} reduces to 
{the form given in \eqref{LL-sec-Lind} with individual} 
secular Lindblad jump operators 
{given in \eqref{Lind-glob-sec} in the main text.}
The Lindblad superoperator {$\L^\text{sec}$ in \eqref{LL}} is thus positivity preserving.

As we see, the secular approximation automatically removes the imaginary parts $B_\al$ 
since they only appear off--diagonal in \eqref{Redfield-matrix} while the diagonal terms $M^\al$ 
with $\D E = \D E'$ are real.

\subsubsection{Global coherent approximation}

Because the secular approximation is losing too much physical information (cf. the main text in Sec.~\ref{sec:L-glob-coh} for the discussion), we developed in \cite{B07-Dimer} the \textit{coherent approximation} as a much less invasive method of restoring the positivity, cf. also \cite{Breuer-CohLind, Kirsanskas+Wacker-CohLindPhenom, Davidovic2020, Nathan-CohLindDerivation}.
The arithmetic mean of the real parts in the off--diagonal terms of $\M^\pm_{\D E, \D E'}$ is now replaced by the geometric mean while the imaginary parts are removed%
\footnote{The replacement is motivated by the properties of 
the negative eigenvalue of $\M^{\pm}_{\D E, \D E'}$ whose sign is proportional to the sign of $G^2 - A^2 - \D B^2 \leq 0$ where $G$ is the geometric mean, $A$ is the arithmetic mean of $f_\pm(\D E)$ and $ f_\pm(\D E')$ (with $G \leq A$) and $\D B = [{B_\pm(\D E) - B_\pm(\D E')}]/{2}$, present in its off--diagonals.
Replacing $A$ with $G$ and setting $\D B = 0$ in \eqref{Redfield-matrix} lifts the negative eigenvalue exactly to zero.}
\begin{align}
  \M^{\pm,\text{coh}}_{\D E, \D E'}  = \Ga
  \begin{pmatrix}
    f_\pm(\D E) & \sqrt{f_\pm(\D E) f_\pm(\D E')} \\
    \sqrt{f_\pm(\D E') f_\pm(\D E)} & f_\pm(\D E')
  \end{pmatrix}
\end{align}
which is the minimal change that shifts the negative eigenvalue of $\M^{\pm}_{\D E, \D E'}$ up to zero%
\footnote{In general, the off--diagonal terms could also get arbitrary phases $\exp[ i (\Phi_{\D E} - \Phi_{\D E'})]$ but this would violate the compatibility with the Redfield matrix \eqref{Redfield-matrix}, at least in some important limiting cases, like $T\ra\infty$ \cite{LL-MTh}.} 
while keeping the diagonal elements of the matrix untouched.  
The diagonal elements decide about the probabilities or forbidden transitions (between populations), therefore they should not be modified. 
In other words, we keep the diagonal (hence the trace) fixed while bringing the determinant to zero.
In consequence, the matrices $\M^{\pm,\text{coh}}_{\D E, \D E'}$ become rank one and the Liouville operator {\eqref{LL}} can be again written in the {reduced form \eqref{LL-sec-Lind}} with now the \textit{coherent} Lindblad jump operators {given in \eqref{Lind-glob-coh}.}

The arithmetic and geometric means mentioned above are equal only%
\footnote{Note that only for $T>0$ is $f$ monotone and the equality implies $\D E = \D E'$. For $T=0$, $f$ is a step function for which many $\D E \neq \D E'$ can satisfy the equality.}
if $f_\pm(\D E) = f_\pm(\D E')$, when the coherent approximation is exact (i.e. identical with the Redfield version). 
This is, for instance, satisfied for all energies in the limit $T\ra\infty$ or for $T=0$ if $\D E$ and $\D E'$ have the same sign (with the chemical potential set to $\EF = 0$). If $\D E$ and $\D E'$ have different signs, the product $f_\pm(\D E) f_\pm(\D E')$ vanishes (at $T=0$) and the result is identical with that of the secular approximation. 
For non--zero temperatures, there holds the standard inequality between the means, 
$[f_\pm(\D E) + f_\pm(\D E')]/2 \geq \sqrt{f_\pm(\D E) f_\pm(\D E')} \geq 0$, 
from which we conclude that the matrix elements $\M^{\pm}_{\D E, \D E'}$ for the coherent approximation are always between those for the exact (Redfield) and the secular case.

{
The coherent approximation is exact (i.e. identical with the Redfield equation) in the limit $T\ra\infty$ or at $T=0$ for processes $\rho \ra L(\D E)\, \rho\, L\s(\D E')$ for which the energy differences $\D E$ and $\D E'$ 
have the same sign (with the chemical potential set to $\EF = 0$). 
If $\D E$ and $\D E'$ have different signs, at $T=0$,
the corresponding terms vanish and 
the result is identical with that of the secular approximation. 
For finite temperatures, the coherent approximation is somewhere between the Redfield and the secular case.
}

{
The action of all global Lindblads on the (non--degenerate) ground state with energy $E_0$ at $T=0$ is zero as a consequence of the factors $f_+(E_i - E_0) \ra \Theta(E_0 - E_i)$ vanishing for $E_i > E_0$ as $T\ra 0$.
Thus, the ground state is automatically stable and decoherence free. 
For degenerate ground states it must be additionally verified whether direct transitions between them (via exchange of particles with the environment) described by the Lindblad operators are possible. 
In our model, all ground states have the same number of particles and therefore no such transitions occur.
}

{
The indices $(\pm,m,s)$ which enumerate the Lindblad operators $L^\pm_{m,s}$ refer to the effective channels between the baths and the system through which the electrons can be exchanged 
and are related to the involved creation and annihilation operators $c^\pm_{m,s}$. 
Although the channels are based on \textit{local} bath--site interactions, the action of the corresponding Lindblad operators is \textit{global} and is expressed in terms of the \textit{global} eigenstates of the system. 
This is related to the fact that the Lindblad operators are derived under the assumption of infinite past evolution (second Markov assumption, limit $t_0 \ra -\infty$ in \eqref{F(t)}) which effectively introduces non--local tunneling possibilities between any bath and system site even if they are not directly connected by the interaction Hamiltonian $H_I$.
}

{
\subsubsection{Local secular and coherent approach}}

{
The corresponding local operators can be obtained formally by taking the limit $J\ra 0$ 
in the global Lindblad operators \eqref{Lind-glob-coh} or \eqref{Lind-glob-sec} (in the secular version, before applying the deltas $\d_{\pm \D E, E_i-E_j}$ separating the terms with different energy differences),
as discussed in  \cite{Global-Local-Sec-2HarmOsc} 
and \cite{GSchaller-Review-NQSys}. 
In the secular case we obtain for each site $m$ and spin $s$ two Lindblad operators,
separately for electron creation ``+'' and annihilation ``-'',
\begin{align} \label{Lind-loc-sec_App}
  \ell^\pm_{m,s,1} &= \sqrt{f_\pm(\eps)}\,c^\pm_{m,s}\,(1-n_{m,\bar s}), \nonumber \\
  \ell^\pm_{m,s,2} &= \sqrt{f_\pm(\eps + U)}\,c^\pm_{m,s}\,n_{m,\bar s},
\end{align} 
for the first electron (e.g. $\ket{0} \lra \ket{\su}$) and for the second electron (e.g. $\ket{\su} \lra \ket{\su\sd}$) in the site $m$ ($\bar s$ stands for the opposite spin to $s$).
For $T=0$ (and $\eps = -U/2$) they reduce to
\begin{align}
  L^+_{m,s,1} &= c^+_{m,s} (1-n_{m,\bar s}), &
  L^+_{m,s,2} &= 0, & \nonumber \\
  L^-_{m,s,1} &= 0, &
  L^-_{m,s,2} &= c^-_{m,s} n_{m,\bar s}, &
\end{align}
which describe the decay of holon ($L^+_{m,s,1} \ket{0}_m = \ket{s}_m$) and doublon ($L^-_{m,s,2} \ket{\su\sd}_m = \ket{\bar s}_m$) states.
}


{
In the coherent case we obtain for each site $m$ and spin $s$ one Lindblad operator which is a coherent sum of the above two contributions, 
separately for electron creation ``+'' and annihilation ``-'',
\begin{align} \label{Lind-loc-coh_App}
  \ell^\pm_{m,s} &= \ell^\pm_{m,s,1} + \ell^\pm_{m,s,2}. 
\end{align}
The local coherent and local secular Lindblads differ only by the action on coherences between different numbers of particles at one site%
\footnote{Which for one site are unphysical but for more sites can potentially play a role (cf. \cite{B07-Dimer}).}.
}

{
For $T \ra \infty$ the local coherent Lindblad operators reduce to 
\begin{equation} \label{L-loc-coh-Tinf}
  L^\pm_{m,s} = \frac{1}{\sqrt{2}} c^\pm_{m,s}
\end{equation}
and are equal to the global coherent Lindblad operators and to the Redfield operators, \eqref{Redfield-matrix}.
This is the advantage of the coherent operators over the secular ones which do not satisfy such compatibility relations. 
}

{
Generally, it might be expected that for short times the local operators should give a better approximation than the global ones because initially the local site--bath couplings induce only local dynamics.
At later times, when correlations between the system and the baths develop and saturate, the global Lindblad operators would become more accurate.
This simplified picture is, however, not quite correct,
for the global coherent Lindblads can, because of their coherence, also well describe localized states.
Only the global secular Lindblads fail to describe localized states properly \cite{LL-MTh}.
Moreover, owing to the quantum mechanical nature of the interactions between the site and the bath, at short timescales, all bath energy levels contribute to the dynamics as there is no strict energy conservation.
In consequence, the short time dynamics should be rather described in a more complicated way, 
using time--dependent Lindblad operators, as discussed in 
\cite{MarkovMemory+Slippage, StayingPositive, Hatrmann+Strunz-RedfieldNonpositivity}.
In \cite{NS+LL+EK-TimedepLindblads} we propose such a method based on an effective time--dependent temperature, $T_\text{eff}(t)$,
which for short times appears high ($T_\text{eff}(t_0^+)\ra\infty$) 
and the Lindblad operators become local \eqref{L-loc-coh-Tinf}.
Only at late times, they converge to the static global form given by \eqref{Lind-glob-coh}.
At low temperatures $T$ the time--dependent Lindblads delocalize rather quickly (together with the increase of entanglement in the system) while at high temperatures $T$ they stay effectively local (because of the strong local decoherence introduced by the baths) forever.
}

\vskip 1em
\section{Relaxation of $N(t)$ {and $H_{JU}(t)$} with global Lindblads} \label{app:Nt-Lind}

{
For the total particle number $N(t)$, 
the global coherent Lindblads satisfy $[L^\pm_{m,\si}, N] = \mp L^\pm_{m,\si}$ and \eqref{Lind-A} leads to
\begin{equation} \label{N-early-glob-app}
  \p_t N(t) = 
  \Ga \sum_{m,\si} \Big( (L^+_{m,\si})\s L^+_{m,\si} - (L^-_{m,\si})\s L^-_{m,\si} \Big).
\end{equation}
}
Inserting the definitions of the global (coherent) Lindblad operators $L^\pm_{m,\si}$ from \eqref{Lind-glob-coh} into \eqref{N-early-glob-app} we arrive at terms containing $\sum_i f_+(E_i-E_j) \ketbra{\chi_i}{\chi_i}$ which can be cast as spectral filters passing only transitions lowering the system energy $H_S$ below the reference value $E_j$ (the Fermi distribution for $T=0$ enforces strictly $E_i<E_j$). 
The formula can be brought to the form of a trace over these filters which effectively counts the possible number of decay channels from the current state ({$\rho(t)$}) to the lower energy states by creation($+$) or annihilation($-$) of one electron at one site.
This number is then very close to twice (because of spin) the difference of the numbers of holon $N_H$ and doublon $N_D$ occupations (defined w.r.t. the local basis) modified slightly by the projection operators (defined w.r.t. the energy basis). It can be shown that the error is of the order of $J/U$ what follows from the structure of the energy sectors $\Sigma_l$ built essentially out of states with fixed numbers of doublons and holons with only small contributions, of the order of $J/U$, from other sectors. 
Since this difference can be written again in terms of the total number of particles, $N_H - N_D = M - N$,
we finally arrive at a closed differential equation for $N(t)$ 
\begin{equation}
  \p_t {N(t)} = \left[2 \Ga + \Ocal{\frac{J}{U}} \right] (M-{N(t)}).
\end{equation}
{
Summarizing, we were able to estimate the error of this approximation by $\Ocal{J/U}$ from the mismatch between the local (w.r.t. $H_U$) and global (w.r.t. $H_S$) eigenstates which is of the order of $J/U$. In practice, the error is even smaller, such that the solution is again exponential 
\begin{equation}
  N(t) \approx M + [ N(0) - M ]\, e^{-2\Ga'\, t}.
\end{equation}  
}

{
Regarding the total energy of the system,
for $T=0$ 
we can show a general energy decay, $dH_{JU}/dt < 0$, in the system%
\footnote{
The inequality is strict only for the secular \cite{spohn1978b, alicki1979a} but almost always true with only tiny deviations for the coherent global Lindblads. 
For coherent Lindblads higher energetic coherences (which do not contribute to the total energy) can decay to the lower energy states (which do contribute) and thus the measured energy can temporarily slightly increase. 
}
towards the low--energy sector of stationary states which decouple from the baths. 
This sector, as discussed above, consists essentially of half--filled sites with possible small contribution of the order of $J/U$ of double and zero occupations.
}

{
However, in the regime of early relaxation, when the system is still far from the low--energy sector, the difference between the action of the local and global Lindblad operators turns out to be relatively small and can be conservatively estimated by $\Ocal{2^M \Ga J}$.
This results in a differential equation similar to \eqref{Ht-local} for the local Lindblads
\begin{equation}
  \p_t H_{JU}(t) = - 2\, \Ga\, H_{JU} + \Ocal{2^M \Ga J}
\end{equation}
with the solution 
\begin{equation}
  H_{JU}(t) = H_{JU}(0)\, e^{-2\,\Ga t} + \Ocal{2^M J}.
\end{equation}
}

{\section{Spin waves and spin current}}

{\subsection{Spin waves} \label{app:Spin-waves}}

{
In order to discuss spin waves,
we choose two spin states $\ket{Z^+} = \ket{\su, ..., \su}$ and $\ket{Z^-} = \ket{\sd, ..., \sd}$ where $\su, \sd$ denote the spin ``up'' and ``down'' along the (arbitrarily chosen) $z$--direction
with the maximal ferromagenetic order $\ev{O_1} = \frac{M}{4}$
and study spin flips around them. 
The total spin in the $z$--direction $\ev{S^z} = \sum_m \ev{S^z_m}$ 
defines then $M+1$ subsectors associated with the expectation values $-\frac{M}{2}, -\frac{M}{2}+1, ..., \frac{M}{2}-1, \frac{M}{2}$. 
In the states $\ket{Z^\pm}$, the spin--$z$ takes the extremal values $\ev{S^z} = \pm \frac{M}{2}$.
}

{
For the investigation of further Heisenberg states,
it is convenient to introduce the local spin--flip, or spin(-$z$) rising and lowering, ladder operators at site $m$
\begin{equation} \label{spin-ladder-ops_App}
  S^+_m = c\s_{m,\su} c_{m,\sd}, \qquad S^-_m = c\s_{m,\sd} c_{m,\su}
\end{equation}
which can be also written as $S_m^\pm = S_m^x \pm i S_m^y$. 
In the spin space 
they satisfy the commutation relations
\begin{align} \label{S_m-CCR}
  [S^+_m, S^-_n] &= 2 S^z_m \d_{m,n}, & [S^\pm_m, S^\pm_n] &= 0
\end{align}
and on--site anti--commutation relations%
\footnote{In general, $\{S^+_m, S^-_n\} \neq \d_{m,n}$ for $m\neq n$. However, on the special background $\ket{Z^\pm}$, the anti--commutation relations satisfy $\{S^+_m, S^-_n\}_{\ket{Z^\pm}} = \d_{m,n}$ while $[S^+_m, S^-_n]_{\ket{Z^\pm}} = \pm \d_{m,n}$.}
\begin{align} \label{S_m-CAR}
  \{S^+_m, S^-_m\} &= 1, & \{S^\pm_m, S^\pm_m\} &= 0.
\end{align}
By applying the Wigner--Jordan transformation \cite{Jordan-Wigner} the excitations can be interpreted as (spinless) fermions%
\footnote{We have on--site anti-commutation $\{S_m^+,S_m^-\} = 1$ and off--site commutation relations $[S_m^+, S_n^-]=0$  for $m\neq n$ which can get ``fermionized`` by the Wigner-Jordan transformation.
However, it makes the new ladder operators non--local and in case of a periodic chain leads to more complex, non--periodic forms of the operators we want to study.
}.
}

{
The Fourier transformed spin wave ladder 
operators \eqref{spin-ladder-ops_App} are defined by 
\begin{equation} 
	\t S^\pm_k = \frac{1}{\sqrt{M}} \sum_m e^{\pm i k m} S^\pm_m
\end{equation}
for the quantized spin wavenumbers $k = 2 \pi s / M$, $s \in \{0, 1, ..., M-1\}$.
Already for the first ``excited'' states $\ket{S^\pm_m} \equiv S^\mp_m \ket{Z^\pm}$ with $\ev{S^z} = \pm (\frac{M}{2}-1)$ an interesting spin dynamics can be observed.
In this subsector, a distinguished role is played by the basis%
\footnote{We use the convention for the order of the fermionic operators $\ket{..., \su \sd, \su, ...} = c\s_{n,\su} c\s_{n,\sd} c\s_{n+1,\su} \ket{..., 0, 0, ...}$.}
of spin waves 
\begin{equation} 
\begin{split}
	\ket{\t S^-_k} &= \t S^+_k \ket{Z^-} = \frac{1}{\sqrt{M}} \sum_m e^{i k m} S^+_m \ket{Z^-} 
	\\
	&= \frac{1}{\sqrt{M}} \big[ 1 \ket{\su, \sd, ..., \sd} + e^{ik} \ket{\sd, \su, \sd, ..., \sd} \\
	&\phantom{= \frac{1}{\sqrt{M}} \big[}+ ...
	+ e^{(N-1)ik} \ket{\sd, ..., \sd, \su} \big], \\
	\ket{\t S^+_k} &= \t S^-_k \ket{Z^+} = \frac{1}{\sqrt{M}} \sum_m e^{i k m} S^-_m \ket{Z^+}.
\end{split}
\end{equation}
Then, the spin wave excitation number, relative to the background state $\ket{Z^\mp}$,
\begin{equation}
    N^{(\mp)}_S = \sum_m S^\pm_m S^\mp_m = \sum_k \t S^\pm_k \t S^\mp_k = M - N^{(\pm)}_S
\end{equation}
determines the number of flipped spins, equal to $S^z \pm M/2$.
Using the identity
$ S^x_n S^x_m + S^y_n S^y_m = \frac{1}{2} ( S^+_n S^-_m + S^-_n S^+_m ) $
the Heisenberg spin--spin interaction $\v S_n \cdot \v S_m$ can be 
split into
the $z$ component $O_z$ describing the spin order along the $z$--axis and the $x$ and $y$ components, cast together in $\Okin$, describing the ``kinetic energy'' of the hopping of the flipped $z$--spins (called in the literature \textit{magnons} \cite{Kittel-Book})
\begin{multline} \label{ES+OS}
\begin{split}
  O_1 &= \frac{1}{2} \sum_m  \left( S^+_{m+1} S^-_m + S^+_{m} S^-_{m+1} \right) + \sum_m S^z_m S^z_{m+1} \\
  &\equiv \Okin + O_z.
\end{split}
\end{multline}
Diagonalized in the spin wave basis, we find the dispersion relation between the wavenumber $k$ and the kinetic energy $\Okin$
\begin{equation}
  \Okin = \sum_k \cos(k) \t S^+_k \t S^-_k \\.
\end{equation}
For the single spin waves, $\ket{\t S_k^\pm}$, containing one flipped spin w.r.t. the fully spin polarized, ferromagenetic background, the order is always $\ev{O_z}_k = \frac{M}{4} - 1$ which together gives
\begin{equation} 
  \ev{O_1}_k = \cos(k) + M/4 - 1
\end{equation}
and the Heisenberg energy becomes
\begin{equation} 
  \ev{H_\Heis}_k = \frac{4J^2}{U} \left( \cos(k) - 1 \right) \leq 0.
\end{equation}
For the double (and higher) spin waves, e.g. $\t S^\pm_k \t S^\pm_{k'} \ket{Z^\mp}$,
the situation is more complex owing to the fact that two different ladder operators $\t S^\pm_k$ and $\t S^\pm_{k'}$ do not act orthogonally to each other (as pure fermionic operators would%
\footnote{The above mentioned Wigner--Jordan transformation would guarantee this property but, unfortunately, on the cost of making the forms of the states more complex and difficult to treat.
}).
For $N \geq 4$  spin waves with one excitation do not form the ground state which contains at least two such ``excitations'' or a linear combination thereof (cf. examples for particular values of $M$ below).
We put forward the hypothesis that the ground state can be always constructed by superimposing several spin waves of which each contributes with the negative energy $H_\Heis$.
We give below a few examples but postpone a general discussion to the future.
}

{\subsection{Spin current} \label{app:Spin-current}}

{Here, we want to derive quantities conserved in the low--energy sector, obtained from their counterparts in the original system. This may be a more complicated way than studying the Heisenberg system itself, but it offers better insight and interpretation of the conservation laws.
In the original system \eqref{H_SJU}, if $U=0$ and the hopping of particles is described only by $H_J$,
the total number of particles $N$ as well as each component of the total spin $\v S$ are conserved:
$[ H_J, N] = 0$, $[ H_J, \v S] = 0$.
There exists also a corresponding conserved particle current \eqref{J_N}, $[H_J, J_N] = 0$, 
and a spin polarized current \eqref{J_S}, $[H_J, \v J_S] = 0$.
The UDL--transformation \eqref{UDL} transforms the original currents to new expressions $\t J_N = W\s J_N W$ and $\t{\v J}_S = W\s \v J_S W$ valid in the low--energy sector.
While there is no particle motion  in the low--energy sector (described by the Heisenberg Hamiltonian), $\t J_N = 0$,  
the spin current is given by
\begin{equation}
  \t {\v J}_{S} = -\frac{1}{U} \left. \{ H_J, \v J_S \} \right|_{\Si_0}.
\end{equation}
From the above it follows $[H_J^2, \{ H_J, \v J_S \}] = 0$ which gives almost the expression
$[H_\Heis, \t{\v J}_S]$ 
except for the projections ($P_0$) onto the lowest energy sector ($\Si_0$). 
However, these turn out to have no effect%
\footnote{In general, $[H_\Heis, I_S^j] = 8 J^2/U \times$
$$ i\sum_m \left[(S^j_{m-1}+S^j_{m+1}) \v S_m \cdot \v S_{m+1} - \v S_m \cdot \v S_{m+1} (S^j_m+S^j_{m+2}) \right] $$
for $j=1,2,3$
vanishes for $M=2, 3$ because of periodicity.
} for $M=2, 3$ and give the conservation laws
\begin{align}
  [H_\Heis, \v S] &= 0, & [H_\Heis, \t{\v J}_S] &= 0, 
\end{align}
which hold only approximately
for $M\geq 4$.
It means that in the low--energy sector, when the hopping of flipped spins is described by $H_\Heis$, 
each component of the total spin $\v S$ 
is conserved together with the related (for $M=2,3$) conserved pure spin current $\t{\v J}_S$, rewritten as a dimensionless operator
\begin{equation}
  \v I_S \equiv - U \t {\v J}_{S} = 2 \sum_m \v S_m \times \v S_{m+1}.
\end{equation}
Its $z$--component can be expressed in terms of the ladder operators $S^\pm_m$  
\begin{equation} \label{I_S_App}
  I_S^z = \sum_m i(S^+_{m+1} S^-_m - S^+_m S^-_{m+1})
\end{equation}
and bears a formal similarity to $J_N$, cf. \eqref{J_N}, as it measures the flow of the flipped spins hopping on the lattice%
\footnote{For $M=2$ this operator is identically zero but can give a local spin current between the two sites when the periodic boundary condition is removed and the sum is reduced to $m=1$ only.}.
In the Fourier space, the spin current takes the simple form
\begin{equation} \label{J_S-MM_App}
  I_S^z = \sum_k 2 \sin(k)\, \t S^+_k \t S^-_k.
\end{equation}
Summarizing, the single spin waves $\ket{\t S^\pm_k}$ are eigenstates of the spin--$z$, $S^z$, order, $O_1$, spin current operator
\begin{equation} \label{I_S-eig_App}
  I_S^z \ket{\t S^\pm_k} = 2 \sin(k) \ket{\t S^\pm_k}
\end{equation}
and of the Heisenberg Hamiltonian $H_\Heis$ 
\begin{equation} \label{Heis-eig_App}
  H_\Heis \ket{\t S^\pm_k} = \frac{4J^2}{U} \left( \cos(k) - 1 \right) \ket{\t S^\pm_k}.
\end{equation}
}

{\subsection{Trimer spin states} \label{app:Trimer-Spin-states}}

{
From the algebraic point of view, the three electron spins of $\frac{1}{2}$  combine to a $S=\frac{3}{2}$ quadruplet and two $S=\frac{1}{2}$ doublets.  The degeneracy of the latter can be broken by  the cyclic lattice shift operator which obtains the physical interpretation of the spin current.
This algebraic structure is the consequence of the $SU(2)$ symmetry combined with the spin frustration on a triangle which eliminates the standard $S=0$ sector and moves the ground state to the  $S=\frac1{2}$ sector.
The spin waves are then pairs of \textit{Goldstone bosons} for each spin current sector with quantized spin wavenumber $k_s$.
The energies of the states with $k_s \neq 0$ satisfy $E(k_s) < E(0)$ thus giving  (degenerate) spin--wave ground states carrying non--zero spin current.
}

{
In the Heisenberg sector where each site is occupied by exactly one electron, four subsectors can be distinguished, relevant from the point of view of spin--waves, categorized by the values of $S^z$, cf. Tab. \ref{tab:spinwaves-alg}:
\begin{enumerate}
  \item $\ket{Z^+} \equiv \ket{\su,\su,\su}$ ($S^z = \frac{3}{2}$),
  \item 3 states $\ket{S^+_s}$ with two spins $\su$ and one $\sd$ ($S^z = \frac{1}{2}$),
  \item 3 states $\ket{S^-_s}$ with two spins $\sd$ and one $\su$ ($S^z = -\frac{1}{2}$),
  \item $\ket{Z^-} \equiv \ket{\sd,\sd,\sd}$ ($S^z = -\frac{3}{2}$).
\end{enumerate}
The two extreme cases with ferromagnetic order ($|S^z| = \frac{3}{2}$, $S = \frac{3}{2}$, $H_S = -\frac{3}{2} U$, $H_J = H_U = 0$, $N_{DH} = 0$, $J_S = 0$) 
are also exact eigenstates of the Hubbard Hamiltonian $H_S$ and
are non--dynamical at $T=0$ for there is no hopping and no exchange with the environment.
The two middle cases with $|S^z| = \frac{1}{2}$ (and $S = \frac{1}{2}$ or $\frac{3}{2}$) are mirror--symmetric to each other so it is sufficient to study only one of them. 
The states can be treated as spin waves with wavenumbers $k = \frac{2\pi}{3} s$ and classified by the eigenvalues of the spin current operator \eqref{I_S-eig}
which for $M=3$ reduces to 
\begin{equation} \label{I_S_M=3}
  I_S^z = \sum_{\ka = \pm} \sum_{s = -1, 0, 1} \ka \sqrt{3}\, s \ket{S^\ka_{s}}\bra{S^\ka_{s}}.
\end{equation}
In this section, we skip the tilde in the spin wave states, as we no more come back to the lattice representation, and enumerate them by the symbol $s \in \{0, \pm\}$ or equivalently an integer number, $s \in \{0, \pm 1\}$, instead of the wavenumber $k$, and the background symbol $\ka = \pm$.
}

{
The three different spin waves of the from \eqref{spin-waves} w.r.t. the background $\ket{Z^-}$ are
\begin{align}
  \ket{S^-_0} &= \frac{1}{\sqrt{3}} \Big[ \ket{\su, \sd, \sd} + \ket{\sd, \su, \sd} + \ket{\sd, \sd, \su} \Big], \nonumber \\
  \ket{S^-_+} &= \frac{1}{\sqrt{3}} \Big[ \ket{\su, \sd, \sd} + e^{ik_0} \ket{\sd, \su, \sd} + e^{-ik_0} \ket{\sd, \sd, \su} \Big], \nonumber \\
  \ket{S^-_-} &= \frac{1}{\sqrt{3}} \Big[ \ket{\su, \sd, \sd} + e^{-ik_0} \ket{\sd, \su, \sd} + e^{ik_0} \ket{\sd, \sd, \su} \Big],
\end{align}
where $k_0 \equiv 2\pi/3$. 
Analogously for the ``inverted'' spin waves $\ket{S^+_s}$, $s\in \{0, \pm 1\}$, which are obtained by flipping all spins (``$\sd$''$\; \lra\; $``$\su$'').
Double waves can be identified with ``inverted'' single waves matched by the identity: $- \sqrt{3} S^\pm_+ S^\pm_- \ket{Z^\mp} = \ket{S^\pm_0} = S^\mp_0 \ket{Z^\pm}$, which means that two spin waves with the wave numbers $k = \pm k_0$ on top of each other are equivalent to the flipped single spin wave with $k = 0$, etc.
}

{
The states $\ket{S^\ka_0}$ are also exact eigenstates of the Hubbard Hamiltonian $H_S$ (which is also true for any number of sites $M$ and spin wavenumber $k=0$ for which always $H_U = H_J = 0$).
It is not surprising as it can be seen as a state with maximal spin $S$ but smaller than maximal projection $|S^z|$ which because of the rotational symmetry of the model needs to have the same properties as the states with maximal $|S^z|$ and ferromagnetic order (indeed, $\ket{S^\ka_0}$ also have ferromagnetic order but with respect to different axes in space). 
}

{
In general, it follows from the structure of the Heisenberg Hamiltonian \eqref{Heis} that the antiferromagnetic order is energetically preferred. 
The exact relation between the Heisenberg energy and the total spin is a coincidence for $M=3$ sites where all site pairs are direct neighbors, implying 
$ S^2 = \frac{9}{4} + 2\, O_1$, cf. \eqref{S2_M=3},
hence, the states with minimal Heisenberg energy have also minimal spin order and spin. 
}

{
Since the trimer consists of an odd number of sites, the perfect antiferromagnetic order is not possible and the spin chain (with the periodic boundary) preferring  the antiferromagnetic order must be frustrated. 
The essential values for the spin states are summarized in Tab. \ref{tab:spinwaves-kin}.
From that perspective the parameters w.r.t. the $z$--axis of the ferromagnetic states $\ket{Z^\ka}$ and $\ket{S^\ka_0}$ are different, however, because of the rotational symmetry which identifies them (both have maximal spin $S=\frac{3}{2}$) their invariant observables, such as the total order and energy, must be indistinguishable. 
While the latter are exact eigenstates of the Hubbard Hamiltonian $H_S$,
all other states $\ket{S^\ka_s}$ are approximate eigenstates of $H_S$ for $U \gg J$, as discussed above in Sec. \ref{sec:Low-Energy}.
}

\subsection{The ground state magnons} \label{app:DH1}

In order to find the energy of the magnon state $\ket{\psi_k}$ with non--trivial $k=\pm k_0$ analytically we act with the system Hamiltonian $H_S$ on the pure spin--wave state $\ket{S_k}$ several times and find out that the operation generates a closed 3--dimensional subspace spanned by $\ket{S_k}, \ket{\xi_k}$ and $\ket{\zeta_k}$ with the latter two consisting of doublon--holon pairs (and one single occupation) in various permutations over the sites. 
The action of the Hamiltonian on these three states can be written as
\begin{equation}
  H \begin{pmatrix} S_k \\ \xi_k \\ \zeta_k \end{pmatrix} = 
  \begin{pmatrix}
     -\frac{3}{2}U & \sqrt{6} J & 0 \\
     \sqrt{6} J & -\frac{1}{2}U & \sqrt{3} J \\
     0 & \sqrt{3} J & -\frac{1}{2}U
  \end{pmatrix}
  \begin{pmatrix} S_k \\ \xi_k \\ \zeta_k \end{pmatrix}
\end{equation}
whose lowest eigenvalue is approximately given by $E_k \approx -\frac{3}{2}U - \frac{6 J^2}{U}$ and the corresponding eigenstate is 
\begin{equation}
  \psi_k \approx \left(1-\frac{3J^2}{U^2}\right) S_k - \frac{\sqrt{6} J}{U} \xi_k + \frac{3 \sqrt{2} J^2}{U^2} \zeta_k.
\end{equation}
These are the ground states of the $M=3$ Hubbard system \eqref{Hubbard}.

{\section{$M=4$} \label{app:M=4}}

{
For $M=4$ the spin wavenumbers are $k \in \{0, \pm \frac{\pi}{2}, \pi\}$ or $k = s\, k_0$ with $s \in \{0, 1, 2, 3\}$, $k_0 = \frac{\pi}{2}$
and the spin order of the single spin wave  $\ket{\t S_k}$ is, cf. \eqref{O_k},
\begin{equation}
  \ev{O_1}_k = \cos(k) \in \{1, 0, -1\}
\end{equation}
and the energy, cf. \eqref{Heis_k}, 
\begin{equation}
  \ev{H_\Heis}_k = \frac{4 J^2}{U} (\cos(k) - 1) \in \left\{0, - \frac{4 J^2}{U}, -\frac{8 J^2}{U}\right\}.
\end{equation}
According to the hypothesis formulated above, only the spin waves with negative energy should contribute to the ground state.
Owing to the symmetry of states for $M=4$ the ground state needs to be built from two spin waves created on the ferromagnetic background.
Indeed, 
the singlet, $S=0$, ground state can be composed as
\begin{equation}
  \ket{\Psi_{\rm af+}} = \frac{1}{\sqrt{3}} (\t S^\pm_2 \t S^\pm_2 - \t S^\pm_1 \t S^\pm_3) \ket{Z^\mp},
\end{equation}
where the spin wave ladder operator $\t S^\pm_2$ with $k=\pi$, corresponding to the most negative energy, acts twice (as in the Fourier space it has no more the ``fermionic'' nature which it has on--site in the lattice space \eqref{S_m-CAR}). 
It is combined with $S^\pm_1$ and $S^\pm_3$ with $k= \pm \frac{\pi}{2}$ which also contribute with negative energy.
The second lowest energy state ($S=1$) takes also a simple form in terms of the spin wave operators%
\footnote{The state must be extra normalized because the action of a product of two spin wave creators, because of their mixed algebra, cf. \eqref{S_m-CCR} and \eqref{S_m-CAR}, does not preserve the norm.}
\begin{equation}
  \ket{\Psi_{\rm af-}} = {\sqrt{2}}\, S^\pm_2 S^\pm_0  \ket{Z^\mp},
\end{equation}
where the wave $k=\pi$ with the largest negative energy is combined with the neutral energy wave $k=0$.
{It is important to mention that the order $O_1$ and energy $H_\Heis$ do not obey a simple linear addition rule for the double and higher spin waves but show nonlinear effects having their origin in the spin--$z$ order part $O_z$, cf. \eqref{ES+OS} (while $\Okin$ behaves additively in the studied examples).}
}

{
The total spin can be written as
\begin{align}
  S^2 &= 
  \frac{3M}{4} + 2 \sum_{m=1}^4 \v S_m \cdot \v S_{m+1}
  + 2 \sum_{m=1}^2 \v S_m \cdot \v S_{m+2} \nonumber \\
  &= 3 + 2\, O_1 + 2\, O_2.
\end{align}
where $O_2$ is the second nearest neighbor spin order.
The nearest neighbor order $O_1 = S^2/2 - 3/2 - O_2$ becomes minimal in the singlet sector with the total spin $S=0$
and for maximal second neighbor order $O_2 = 2 \frac{1}{4} = \frac{1}{2}$ (which is the maximal theoretical value for two spins $\frac{1}{2}$).
The minimal order is then realized by the ground state $\ket{\Psi_{\rm af+}}$ and takes the value
$ \ev{O_1}_{\rm af+} = - 2 $
which corresponds to the Heisenberg energy \eqref{Heis}
\begin{equation}
  \ev{H_\Heis}_{\rm af+} = - \frac{12 J^2}{U}.
\end{equation}
For $\ket{\psi_{\rm af-}}$, the nearest neighbor order $O_1$ becomes minimal with the total spin $S=1$
and the maximal second neighbor order $O_2 = \frac{1}{2}$
and takes the value
$ \ev{O_1}_{\rm af-} = - 1 $
which corresponds to the Heisenberg energy \eqref{Heis}
\begin{equation}
  \ev{H_\Heis}_{\rm af-} = - \frac{8 J^2}{U}.
\end{equation}
}

\section{The UDL method} \label{app:UDL}

The popular Schrieffer–Wolff method \cite{Schrieffer-Wolff} has the disadvantage of being implicit and requires solving commutator equations by which it makes difficult to control the corrections order by order. 
Therefore, we propose an alternative method which enables to construct the low--energy Hamiltonian in a fully controlled way.
It is based on the UDL decomposition (or, equivalently, LDU decomposition) of block matrices
\begin{equation} \label{UDL-ABCD}
  \begin{pmatrix}
    A & B \\ C & D
  \end{pmatrix}
  =
  \begin{pmatrix}
    \1 & B D^{-1} \\ 0 & \1
  \end{pmatrix}
  \begin{pmatrix}
    A - B D^{-1} C & 0 \\ 0 & D
  \end{pmatrix}
  \begin{pmatrix}
    \1 & 0 \\ D^{-1} C & \1
  \end{pmatrix}
\end{equation}
which is applied to the system Hamiltonian $H_S$ decomposed into blocks with zero and non--zero doublon--holon pair numbers by means of the projectors $P_0$ and $P_+$, respectively, 
\begin{equation}
  H_S = \begin{pmatrix}
    P_0 H_S P_0 & P_0 H_S P_+ \\ P_+ H_S P_0 & P_+ H_S P_+
  \end{pmatrix}.
\end{equation}
We find
\begin{align}
  P_0 H_S P_0 &= 0, & P_+ H_S P_0 &= P_1 H_J P_0, \nonumber \\ 
  P_1 H_S P_1 &= U P_1 , & P_0 H_S P_+ &= P_0 H_J P_1. 
\end{align}
with $P_1$ being the projection onto exactly one doublon--holon pair. 
This gives the inverse, $P_1 H_U^{-1} P_1 = U^{-1} P_1$, 
required in \eqref{UDL-ABCD} and leads to the UDL decomposition
\begin{align}
  H_S &= L \t H_S R
\end{align}
with 
\begin{align}
  \t H_S &= 
  \begin{pmatrix}
    - \frac{1}{U} P_0 H_J P_1 H_J P_0 & 0 \\ 0 & P_+ H_S P_+
  \end{pmatrix}.
\end{align}
The triangular matrices $L$ and $R$ can be further anti--symmetrized to $L'$ and $R'$ which are closer to unitarity (hence the name \textit{modified UDL decomposition}), $L'\s L' \approx R'\s R' \approx \1$, and almost inverse to each other, 
\begin{align}
  L' &= \begin{pmatrix}
         \1 & \frac{1}{U} H_J \\ -\frac{1}{U} H_J & \1
       \end{pmatrix} + \Ocal{\frac{J^2}{U^2}}
  = R'\s,
\end{align}
and via $ H_S = L' \t H_S' R' $ generate an almost diagonal 
\begin{align}
  \t H_S' &= \begin{pmatrix}
            -\frac{1}{U} H_J^2 & 0 \\
            0 & U 
          \end{pmatrix} + \Ocal{\frac{J^3}{U^2}}.
\end{align}
This leads to the unitary transformation of the Hamiltonian $H_S$ and of the states $\ket{\psi_k}$ {given by} \eqref{UDL}
with a unitary $W \approx L' \approx R'\s = \1 + \Ocal{J/U}$.
The left upper block in $\t H_S'$ acts in the lowest energy sector and can be evaluated to the Heisenberg Hamiltonian \cite{cleveland1976a}, cf. \eqref{Heis},
 \begin{equation}
  P_0 \t H_S' P_0 = H_0 +
  \frac{4 J^2}{U} \sum_{\<m,n\>} \left( \v S_m \cdot \v S_n - \frac{1}{4} \right) 
  \equiv H_0 + H_\Heis 
\end{equation}
In the main text, we skip the prime sign in $\t H_S'$.

\bibliography{qdots}

\end{document}